\documentclass[longauth]{aa}
\usepackage[varg]{txfonts}
\usepackage{natbib}
\usepackage{ae,aecompl}
\usepackage{graphicx}   
\usepackage{amsmath}    
\usepackage{amssymb}    
\usepackage{siunitx}    
\DeclareSIUnit{\degree}{deg} 
\DeclareSIUnit{\arcmin}{arcmin} 
\DeclareSIUnit{\arcsec}{arcsec} 
\usepackage{color,verbatim,url}
\usepackage{tabularx}
\usepackage{booktabs}
\usepackage{pdflscape}

\usepackage{minitoc}
\usepackage{hyperref}

\usepackage{breqn}

\definecolor{pink}{rgb}{0.858, 0.188, 0.478}
\definecolor{purple}{RGB}{76, 0,153}

\newcommand{\ie}{{\rm i.e.}}
\newcommand{\eg}{{\rm e.g.}}

\newcommand{\ssim}{{\sim}}
\newcommand{\stimes}{{\times}}
\newcommand{\utoi}{{ugri}}
\newcommand{\utoz}{{ugriZ}}
\newcommand{\utoy}{{ugriZY}}
\newcommand{\utoj}{{ugriZY\!J}}
\newcommand{\utoh}{{ugriZY\!J\!H}}
\newcommand{\utok}{{ugriZY\!J\!H\!K_{\rm s}}}
\newcommand{\gtok}{{griZY\!J\!H\!K_{\rm s}}}
\newcommand{\ztok}{{ZY\!J\!H\!K_{\rm s}}}
\newcommand{\ytok}{{Y\!J\!H\!K_{\rm s}}}

\newcommand{\bbf}{}

\begin{document}

\title{KiDS+VIKING-450: A new combined optical \& near-IR dataset for cosmology and astrophysics}
\titlerunning{KiDS+VIKING-450}

\author{ Angus H. Wright\inst{1,2} 
\and Hendrik Hildebrandt\inst{1,2} 
\and Konrad Kuijken\inst{3} 
\and Thomas Erben\inst{1} 
\and Robert Blake\inst{4} 
\and Hugo Buddelmeijer\inst{3} 
\and Ami Choi\inst{5} 
\and Nicholas Cross\inst{4} 
\and Jelte T.A. de Jong\inst{6} 
\and Alastair Edge\inst{7} 
\and Carlos Gonzalez-Fernandez\inst{8} 
\and Eduardo Gonz\'alez Solares\inst{8} 
\and Aniello Grado\inst{9} 
\and Catherine Heymans\inst{4,1} 
\and Mike Irwin\inst{8} 
\and Aybuke Kupcu Yoldas\inst{8} 
\and James R. Lewis\inst{8} 
\and Robert G. Mann\inst{4} 
\and Nicola Napolitano\inst{10} 
\and Mario Radovich\inst{11} 
\and Peter Schneider\inst{2} 
\and Crist\'obal Sif\'on\inst{12}
\and William Sutherland\inst{13} 
\and Eckhard Sutorius\inst{4} 
\and Gijs A. Verdoes Kleijn\inst{6} }
\authorrunning{The KiDS Collaboration}
\institute{
Astronomisches Institut, Ruhr-Universit\"at Bochum, Universit\"atsstr. 150, 44801 Bochum, Germany\\\email{awright@astro.rub.de} \and
Argelander-Institut f\"ur Astronomie, Auf dem H\"ugel 71, 53121 Bonn, Germany \and 
Leiden Observatory, Leiden University, Niels Bohrweg 2, 2333 CA Leiden, the Netherlands \and 
Institute for Astronomy, University of Edinburgh, Royal Observatory, Blackford Hill, Edinburgh EH9 3HJ, UK \and 
Center for Cosmology and AstroParticle Physics, The Ohio State University, 191 West Woodruff Avenue, Columbus, OH 43210, USA \and 
Kapteyn Astronomical Institute, University of Groningen, PO Box 800, 9700 AV Groningen, the Netherlands \and 
Centre for Extragalactic Astronomy, Department of Physics, Durham University, South Road, Durham, DH1 3LE, UK  \and 
Institute of Astronomy, University of Cambridge, Madingley Road, Cambridge CB3 0HA, UK \and 
INAF -- Osservatorio Astronomico di Capodimonte, Via Moiariello 16, 80131 Napoli, Italy \and 
School of Physics and Astronomy, Sun Yat-sen University, Guangzhou 519082, Zhuhai Campus, P.R. China \and 
INAF - Osservatorio Astronomico di Padova, via dell'Osservatorio 5, 35122 Padova, Italy \and 
Department of Astrophysical Sciences, Peyton Hall, Princeton University, Princeton, NJ 08544, USA \and 
School of Physics and Astronomy, Queen Mary University of London, Mile End Road, London E1 4NS, UK 
}

\date{Released 14/12/2018}


\graphicspath{{./figures/}}

%
%

\abstract{We present the curation and verification of a new combined optical and near
  infrared dataset for cosmology and astrophysics, derived from the
  combination of $\utoi$-band imaging from the Kilo Degree Survey (KiDS) and
  $\ztok$-band imaging from the VISTA Kilo degree Infrared Galaxy (VIKING)
  survey. This dataset is unrivaled in cosmological imaging surveys due to its
  combination of area ($458$ deg$^2$ before masking), depth ($r\le25$), and
  wavelength coverage ($\utok$). The combination of survey depth, area, and
  (most importantly) wavelength coverage allows significant reductions in
  systematic uncertainties (\ie\ reductions of between 10 and 60\% in bias,
  outlier rate, and scatter) in photometric-to-spectroscopic redshift
  comparisons, compared to the optical-only case at photo-$z$ above $0.7$. 
  The complementarity between our optical and NIR surveys means that over
  $80\%$ of our sources, across all photo-$z$, have significant detections
  (i.e. not upper limits) in our $8$ reddest bands. We derive photometry,
  photo-$z$, and stellar masses for all sources in the survey, and verify these
  data products against existing spectroscopic galaxy samples.  We demonstrate
  the fidelity of our higher-level data products by constructing the survey
  stellar mass functions in 8 volume-complete redshift bins. We find that these
  photometrically derived mass functions provide excellent agreement with
  previous mass evolution studies derived using spectroscopic surveys.
  The primary data products presented in this paper are publicly available at 
  \url{http://kids.strw.leidenuniv.nl/}.
  }


\maketitle

\section{Introduction}
\label{sec:intro}
%
Over the last decade observational cosmological estimates have 
become increasingly restricted by systematic, rather than random, 
uncertainties \citep{hildebrandt/etal:2016,hildebrandt/etal:2018,
troxel/etal:2018,hikage/etal:2018,planck/cosmo:2018}. In particular, estimates made
using large photometric samples of galaxies, such as those utilising weak
gravitational lensing \citep[\eg][]{bacon/etal:2000,
vanwaerbeke/etal:2000,wittman/etal:2000,rhodes/etal:2001},
have moved closer to a regime where increasing sample sizes alone are unlikely
to cause a significant improvement in estimate
constraints \citep{becker/etal:2016,jee/etal:2016,
hildebrandt/etal:2017,troxel/etal:2018,hildebrandt/etal:2018}.
Instead, quantification and reduction of systematic biases is becoming increasingly
important, and more frequently is the dominating source of uncertainty in
cosmological inference \citep{mandelbaum:2018}.

One such systematic limitation/bias for many methods of observational
cosmological inference \citep[and indeed one that frequently dominates the systematic 
uncertainty budget; ][]{hildebrandt/etal:2016,hikage/etal:2018} is also one of the most fundamental: that of estimation
of source positions in 3-dimensional space. Specifically, localisation of 
galaxies along the line-of-sight (i.e. distance) axis is of particular importance. 
This localisation is primarily achieved through relatively low-precision
photometric based methods, referred to as photometric redshift or photo-$z$ \citep[see ][for a summary of 
photo-$z$ methods]{hildebrandt/etal:2010}.

One method for deriving photo-$z$ estimates involves finding the model galaxy spectrum, from a 
sample of representative spectrum templates, which best fits the observed galaxy 
flux in a series of wavelength bands \citep{ilbert/etal:2006,benitez:2000,brammer/etal:2008,bolzonella/etal:2000}. 
Such estimates are typically restricted 
by the quality of the input photometry, the intrinsic redshift distribution of the 
source galaxy sample, and the degeneracy between various galaxy spectrum models 
as a function of galaxy redshift \citep{ilbert/etal:2009,laigle/etal:2019}. 
In each of these cases, however, additional information 
can lead to significant benefits in the photo-$z$ estimation process. 

For cosmological inference, a weak lensing survey needs to provide reliable
galaxy shapes \citep[see, \eg, ][]{massey/etal:2007a} and redshift estimates for a
statistically representative sample of galaxies over cosmologically significant
redshifts \citep[see, \eg,
][]{hildebrandt/etal:2018,troxel/etal:2018,hikage/etal:2018}.  For this purpose
there are, therefore, three main properties that determine any weak lensing
survey's cosmological sensitivity: survey area, survey depth, and wavelength
coverage. {\bbf These properties all contribute to both the statistical and
systematic uncertainty on cosmological inference.  The first two statistics,
area and depth, typically govern the raw number of sources (in a given redshift
interval) that can be used for inference; a primary contributor to the
statistical uncertainty on cosmic shear cosmological inference, and so a key
consideration in weak lensing survey design
\citep{dejong/etal:2015,abbott/etal:2015,aihara/etal:2018}.  Similarly, and discussed
in more depth below, the wavelength information dictates the limiting photo-$z$
accessible for tomographic binning \citep{hildebrandt/etal:2018}, which is a
driving factor in the final signal-to-noise of a cosmic shear estimate. As such
wavelength coverage also contributes non-negligibly to the statistical
uncertainty of cosmic shear cosmological estimates.  On the systematic effects
side, area and depth aid in the constraint of many systematics parameters
directly, not the least of which is the intrinsic alignment signal
\citep{joachimi/etal:2015}. Deeper data allows more galaxy satellites to be
observed, and a better constraint on the galaxy-galaxy intrinsic alignment
signal to be made. The wavelength baseline} is also a primary driving factor in
determining the systematic uncertainty on any cosmological inference, primarily
because it dictates the signal-to-noise of the shear signal of individual
cosmic shear sources.  

The reason wavelength information is of significant importance in cosmological 
inference is many-fold. However we focus on two main reasons, for demonstration: namely  
reduction of photo-$z$ bias and influence over the redshift baseline. We discuss both of these 
below. 

The importance of wavelength information in the reduction of photo-$z$ bias is driven mostly 
by the degeneracy between galaxy spectrum models. Even with 
perfect input photometry, there exist degeneracies between galaxy spectrum models at 
different redshifts over finite wavelength intervals \citep[see Figure 1 of ][ for a nice demonstration]{buchs/etal:2019}. 
These degeneracies can lead to 
considerable biases in the source photo-$z$ distribution, {\bbf when sources are systematically assigned to 
incorrect parts of redshift-space. Moreover, these effects are increasingly problematic 
as source photometry becomes noisier and the wavelength baseline becomes shorter. Naturally,} the only way 
to break such degeneracies is by utilising photometry for these sources that extends 
beyond the wavelength range wherein the degeneracy exists. As such, longer wavelength baselines 
are fundamental to the breaking of model degeneracies, and therefore to reducing the {\bbf systematic incorrect 
assignment of photo-$z$. Such incorrect assignment can lead to considerable bias in estimated redshift 
distributions \citep{hildebrandt/etal:2018} which limits cosmological inference, over any given source redshift baseline. }

Furthermore, wavelength information is the primary factor which determines the useful redshift baseline 
over which cosmological inference can be performed. In particular, photometry that extends 
redward of the optical bands is essential for the accurate estimation of photo-$z$ beyond a redshift 
of $z\sim1$ (for typical ground-based photometric surveys). This intermediate- to high-redshift information 
is of particular importance to weak lensing cosmological inference, as higher-redshift sources carry 
considerably more signal-to-noise than their lower-redshift counterparts. 
This increased signal is critical in the quantification of systematic bias as it allows them to be 
explored with reduced stacking of sources (e.g. with finer bins containing more homogeneous samples of 
galaxies), which can alleviate additional biases. 

To date, the largest joint optical and near infrared (NIR) dataset for cosmology was a combined 
Dark Energy Survey (DES) + VISTA Hemisphere Survey (VHS) analysis of the DES Science Verification region, 
covering $\ssim 150$ deg$^2$ and spanning the $\gtok$ bands \citep{banerji/etal:2015}. 
In this paper we present the integration of two European Southern Observatory
(ESO) public surveys; the VISTA Kilo degree INfrared Galaxy
\citep[VIKING;][]{edge/etal:2013,venemans/etal:2015} survey, probing the NIR
wavelengths ($8000-24000$\AA), and the Kilo Degree Survey
\citep[KiDS;][]{kuijken/etal:2015,dejong/etal:2015}, probing optical
wavelengths ($3000-9000$\AA). These combined data represent a significant step forward 
from the previous state-of-the-art, primarily due to the increase in combined survey area 
and optimal matching between the two surveys depth (see Sect.~\ref{sec:data}). 

This extension of the wavelength baseline brings with it considerable benefits,
particularly for cosmic shear analyses. \cite{hildebrandt/etal:2017} presented
cosmological inference from cosmic shear using 450 square degrees of KiDS imaging (referred to as 
KiDS-450), measuring the matter clustering parameter ($\sigma_8$) and 
matter density parameter ($\Omega_{\rm m}$), which are typically parameterised jointly as
$S_8=\sigma_8\sqrt{\Omega_{\rm m}/0.3}$, to a relative
uncertainty of $\ssim5\%$; an error whose budget was limited essentially
equally by systematic and random uncertainties. As a result, we expect that the
final KiDS dataset, spanning $1350$ deg$^2$, will in fact be systematics
limited in its cosmological estimates (as random uncertainties should downscale 
by a factor of roughly $\sqrt{3}$). Moreover, constraint and reduction of
systematic effects will become of increasing importance in the next few years,
and indeed into the next decade with the initiation of large survey programs
such as Euclid \citep[see, e.g.,][]{amendola/etal:2018}. Using the combined dataset 
presented here enables us to make considerable progress regarding the challenge of
reducing systematics, and that in doing so enables us to perform an
updated cosmic shear analysis which better constrains systematic uncertainties and
enables the use of higher-redshift sources \citep{hildebrandt/etal:2018}.  

Importantly, this dataset is not only useful for cosmological studies. The
additional information provided by the near-IR allows better constraint of fundamental 
galaxy parameters such as stellar mass and star formation rates, which enable the 
construction of useful samples for galaxy evolution and astrophysics studies. 
For example, recent use of near-IR data in preselection of ultra-compact massive  
galaxy candidates \citep{tortora/etal:2018} has allowed the spectroscopic confirmation of 
the largest sample of UCMGs to date. 

As such, this work focusses on the description and verification of the joint
KiDS+VIKING photometric dataset, and on the derivation of higher-level data
products which are of interest both for weak-lensing cosmological analyses and
non-cosmological science use-cases. The KiDS optical and VIKING NIR data and
their reduction are described in Sect.~\ref{sec:data}. The multi-band
photometry and estimation of photo-$z$ are covered in Sect.~\ref{sec: photo}.
Model fitting to the broadband galaxy spectral energy distributions is given in
Sect.~\ref{sec: masses}, as is the exploration of stellar mass estimates from
these fits. We compare the resulting stellar mass function for our 
dataset to previous works in Sect.~\ref{sec: mass function}. The paper is
summarised in Sect.~\ref{sec:summary}. The primary data products described in this 
paper are made publicly available at \url{http://kids.strw.leidenuniv.nl/}. 


\section{Dataset and reduction}
\label{sec:data}
%
%
In this section we describe the KiDS optical (Sect.~\ref{sec:KiDS}) and VIKING
NIR (Sect.~\ref{sec:VIKING}) imaging that is used in this study. KiDS and
VIKING are {\bbf partner} surveys that will both observe two contiguous patches of sky
in the Galactic North and South, covering a combined area of over 1350 square
degrees \citep{arnaboldi/etal:2007,dejong/etal:2015,dejong/etal:2017}.
Observations for KiDS are ongoing, and so joint analysis of KiDS+VIKING is
currently limited to the footprint of the third KiDS Data Release
\citep{dejong/etal:2017}. 

\begin{figure*}
\includegraphics[width=\textwidth]{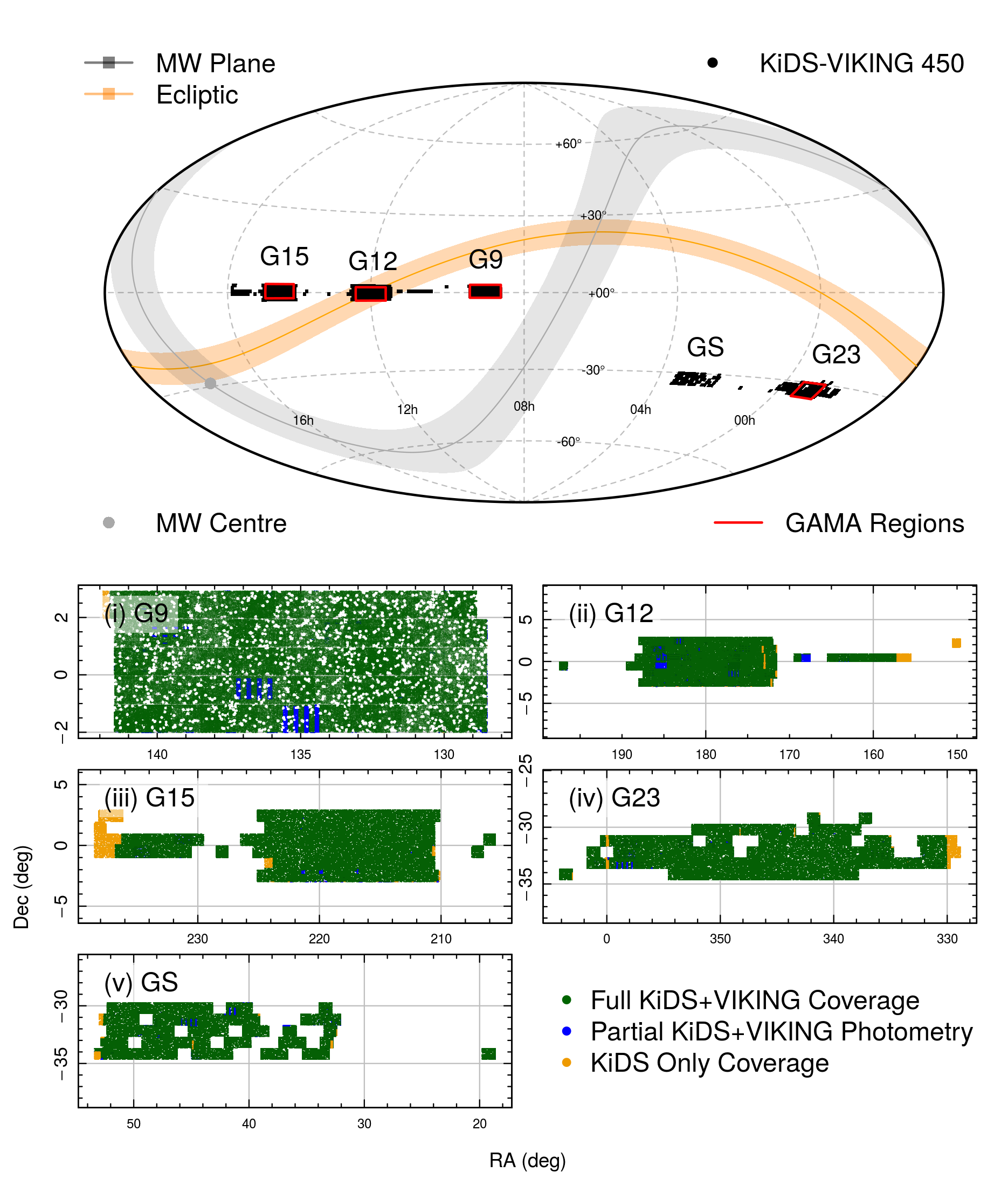}
\caption{\label{fig:footprint} The footprint of the post-masking KiDS-450
  dataset. {\em Top: } the distribution of the KiDS-450 fields on-sky, relative
  to the Ecliptic and Galactic planes. The Galactic plane is plotted with a
  width of 20 degrees, which roughly traces the observed width of the Galaxy
  thick disk. {\em Lower panels: } each of the individually named KiDS-450
  patches (on their own scale). The points in each patch show the distribution
  of KiDS-450 photometric sources that remain after applying the bright-star
  mask. Points are coloured according to their overall observational coverage:
  green points have full KiDS+VIKING optical and NIR coverage, blue points have
  full KiDS optical coverage but only partial VIKING NIR coverage, and orange
  points have KiDS optical coverage only. As such the green and blue data show
  the footprint of the full KiDS+VIKING-450 (KV450) sample. }
\end{figure*}

The footprint of the post-masking KiDS-450 dataset presented in
\cite{hildebrandt/etal:2017} is shown in
Fig.~\ref{fig:footprint} both on-sky and split into each of the KiDS `patches'
(where each patch contains one of the $5$ $\sim$contiguous portions of the
KiDS-450 footprint). These individual patches divide the KiDS-450
survey area into five sections of (roughly) contiguous data on-sky, centering
primarily on fields observed by the Galaxy And Mass Assembly
\citep[GAMA,][]{driver/etal:2011} redshift survey.  The geometry of each 
patch can be seen in the individual panels of
Fig.~\ref{fig:footprint}, and are named by the GAMA field on which they are
focused.  The exception is the GS patch, which has no corresponding GAMA field;
we nonetheless maintain the naming convention for convenience. Note, though,
that future KiDS observations will close the gaps both within and between the
patches, and lead to the creation of a contiguous $\ssim10~{\rm
deg}\stimes75~{\rm deg}$ stripe in both the Galactic North and South.
Observations of these contiguous stripes in VIKING have already been completed. 

The full (unmasked) KiDS-450 dataset consists of $\ssim49$ million non-unique
Galactic and extragalactic sources distributed over $454$ overlapping $\ssim1$
deg$^2$ pointings on sky (see Sect.~\ref{sec:KiDS}).  This reduces to
$\ssim33.9$ million unique mostly-extragalactic sources after applying masking
of stars and removing duplicated data, distributed over $\ssim360$ deg$^2$.
These unique post-masking sources are shown in Fig.~\ref{fig:footprint},
and so the masking around bright stars, for example, can be seen as small
circular gaps within the patches. Each source is coloured by its observational
coverage statistics: those with full photometric KiDS+VIKING observational
coverage are shown in green, those with full KiDS observational coverage but
only partial VIKING observational coverage are shown in blue, and those with
only KiDS observational coverage are shown in orange. {\bbf We define the combined
KiDS+VIKING-450 sample (hereafter KV450) as those KiDS-450 sources which have
overlapping VIKING imaging (i.e. the green sources in
Fig.~\ref{fig:footprint}). Masking the regions with missing near-IR coverage (i.e.
the orange and blue data in Fig.~\ref{fig:footprint}), the full KV450 footprint consists
of $447$ overlapping pointings\footnote{There are a number of pointings, particularly at the
survey edges, which have only slight overlap between KiDS and VIKING. This causes 
the overall loss in area ($\ssim19$ deg$^2$) to be somewhat larger than the
loss of only $7$ full pointings would suggest.}, covering $\ssim341$ deg$^2$,
and consists of $\ssim31.9$ million unique mostly-extragalactic sources. }


\begin{figure*}
\includegraphics[width=\textwidth]{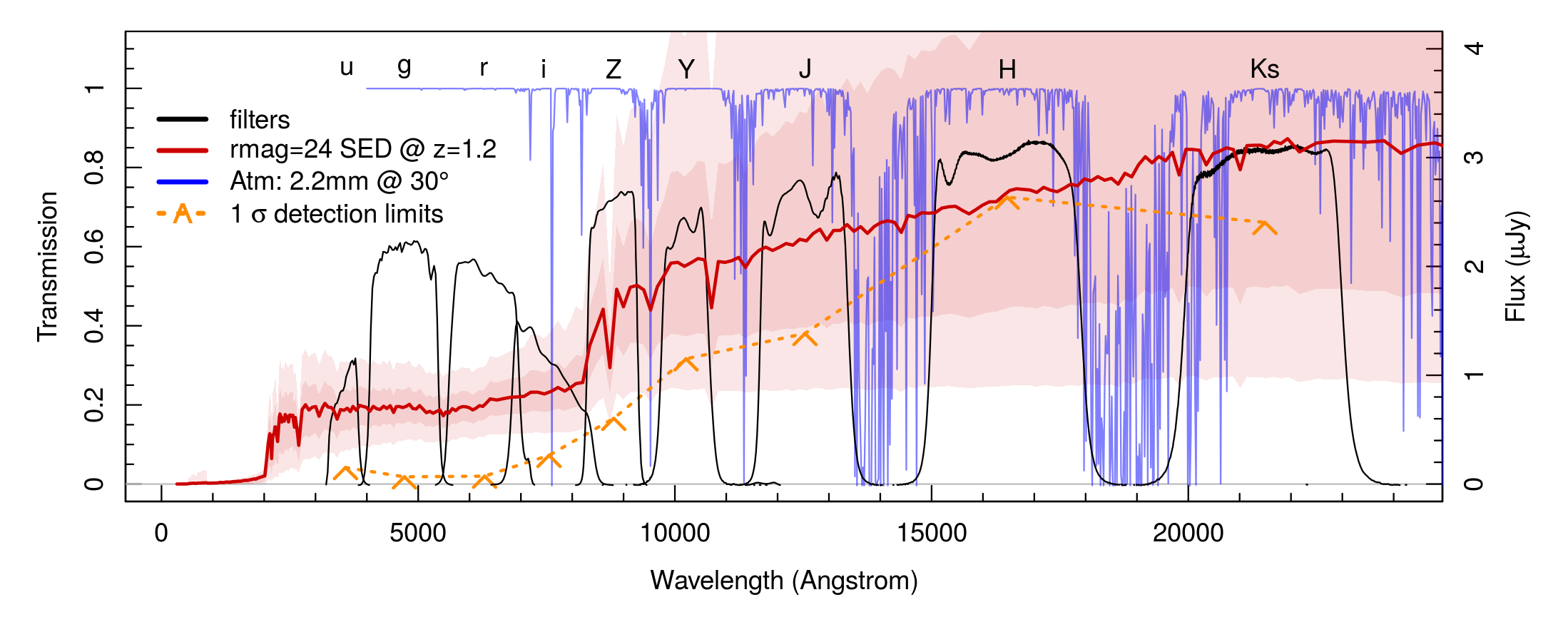}
  \caption{\label{fig:filters} The individual photometric filters (black) that make up
  the KV450 dataset. Each filter curve is shown as an overall transmission
  spectrum incorporating mirror, detector, and filter effects. We also show the
  typical transmission spectrum of the atmosphere at Paranal (blue) for modest
  values of precipitable water vapor (2.2mm) and zenith angle (30 degrees). In
  addition, we also show the {\bbf median {\sc Le Phare} spectrum of all KV450 
  galaxies with photometric redshift $Z_B=1.2$ and magnitude $r\sim24$ (red). The 
  $68^{\rm th}$ and $99^{\rm th}$ percentiles of these models are also shown as shaded  
  red regions. The $1\sigma$ detection limits of each band 
  (orange chevrons and dotted line, derived from the values in \protect
  Table~\ref{tab: detection thres}) are also shown, for reference. These model spectra demonstrate }
  the complementarity of the KiDS \& VIKING surveys; {\bbf a typical galaxy at the furthest and faintest 
  end of our analysis is still detected in all bands. It also demonstrates the main
  benefit of having NIR imaging within this dataset, in that it allows much more
  accurate constraint of photometric redshifts for $(4000$\AA$)$ Balmer-break
  galaxies at redshifts $z \gtrsim 1$.}}
\end{figure*}

\begin{table}
  \caption{Magnitude limits and typical seeing values for each of the KiDS+VIKING photometric bands. 
  Values reproduced from \protect \cite{dejong/etal:2017,venemans/etal:2015}.}\label{tab: detection thres}
  \begin{tabular}{c|cccc}
    Band & $\lambda_{\rm cen}$ & Exp.time (s) & Mag Limit & PSF FWHM \\
         & (\AA) & (s) & $(2^{\prime\prime}\,5\sigma\,{\rm AB})$ & ($\prime\prime$) \\
    \hline
     $u $           & 3550  & 1000  & 24.2  & 1.0  \\
     $g $           & 4775  & 900   & 25.1  & 0.9  \\
     $r $           & 6230  & 1800  & 25.0  & 0.7  \\
     $i $           & 7630  & 1200  & 23.6  & 0.8  \\
     $Z $           & 8770  & 480   & 22.7  & 1.0  \\
     $Y $           & 10200 & 400   & 22.0  & 1.0  \\
     $J $           & 12520 & 400   & 21.8  & 0.9  \\
     $H $           & 16450 & 300   & 21.1  & 1.0  \\
     $K_{\rm s}$    & 21470 & 480   & 21.2  & 0.9  \\
    \hline 
  \end{tabular}
\end{table}

\subsection{KiDS-450 optical data}
\label{sec:KiDS}
%
%
The data reduction for the KiDS-450 $\utoi$-band survey data is described in detail 
in \cite{hildebrandt/etal:2017} and \cite{dejong/etal:2017}, which we briefly 
summarise here.  
As stated previously, the full optical dataset consists of $454$ distinct
$\ssim1~{\rm deg}^2$ pointings of the OmegaCAM, which is mounted at the
Cassegrain focus of ESO's VLT Survey Telescope (VST) on Cerro Paranal, Chile.
Images in the $\utoi$-bands are available for all of these pointings, with
exposure times of $15-30$ minutes and $5\sigma$ limiting magnitudes of
$23.8-25.1$; precise values are given in \cite{dejong/etal:2017} and are
reproduced here in Table~\ref{tab: detection thres}. The filter transmission
curves for these four optical bands are shown in Fig.~\ref{fig:filters}, along
with the atmospheric transmission typical to observations at Paranal.  

The optical data for KV450 are reduced using the same reduction pipelines as in
KiDS-450.  Specifically, the {\sc AstroWISE} \citep{valentijn/etal:2007}
pipeline is used for reducing the $\utoi$-band images and measuring multi-band
photometry for all sources. Independently, the {\sc THELI}
\citep{erben/etal:2005,schirmer:2013} pipeline performs an additional reduction
of the $r$-band data, which is used for cross-validating the {\sc AstroWISE}
reduction and for performing shape measurements for weak lensing analyses. 

The only difference between the KiDS-450 and KV450 optical datasets is that the
KV450 optical reduction incorporates an updated photometric calibration.
KiDS-450 invoked only relative calibration across the $\utoi$-bands with
stellar-locus-regression \citep[SLR,][]{high/etal:2009}. The absolute
calibration of these data was reliant on nightly standard star observations and the 
overlap between u- and r-band tiles to homogenize the photometry. 
Since the publication of \cite{hildebrandt/etal:2017}, the first data
release from the European Space Agency's {\em Gaia} mission has been made
available \citep{GaiaDR1}. Gaia offers a sufficiently homogeneous,
well-calibrated anchor that can be used to greatly improve this absolute
calibration. The calibration procedure is described in \cite{dejong/etal:2017}
and all optical data used here are absolutely calibrated in this way.

\subsection{VIKING infrared data}
\label{sec:VIKING}
%
%
VIKING is an imaging survey conducted with the Visible and InfraRed CAMera
(VIRCAM) on ESO's 4m VISTA telescope. The KiDS and VIKING surveys were designed 
together, with the specific purpose of providing well-matched optical and
NIR data for $\ssim1350$ square degrees of sky in the Galactic North and South.
As such, the surveys share an almost identical footprint on-sky, with minor
differences being introduced due to differences in the camera field of view and
observation strategy. VIKING surveys these fields in five NIR bands
($\ztok$), whose filter transmission curves are shown in 
Fig.~\ref{fig:filters}, and total exposure times in each band are chosen such
that the depths of KiDS and VIKING are complementary. 

A detailed description of the VIKING survey design and observation strategy can
be found in \cite{edge/etal:2013} and \cite{venemans/etal:2015}. Briefly, VIRCAM consists of 16
individual HgCdTe detectors, each with a $0.2\stimes0.2$ square degree angular size, but which
jointly span a $\ssim1.2$ square degree field of view, thus leaving considerable
gaps between each detector.  Observations made by VIRCAM for VIKING therefore
implement a complex dither pattern which is able to fill in the detector gaps
while also performing jittered observations to enable reliable estimation of
complex NIR backgrounds and sampling of data across detector defects. The
observation strategy thus involves taking multiple exposures with small
(i.e. much less than detector width) jitter steps, taken in quick succession, 
which are then stacked together to create a `paw-print'.  The
stacked paw-print still has large gaps between the 16 detectors, and so a
dither pattern of 6 stacked paw-prints is required in order to create a
contiguous $\ssim 1.5$ square degree image, called a tile. The reduction of the
data, and the production of these individual data products (reduced exposures,
stacked paw-prints, and completed tiles), is carried out by the Cambridge
Astronomy Survey Unit \citep[CASU,][]{gonzalez-fernandez/etal:2018,lewis/etal:2010}. 
These reduced data are then transferred to the Edinburgh Royal Observatory Wide
Field Astronomy Unit VISTA Science Archive \citep[WFAU
VSA,][]{Irwin/etal:2004,hambly/etal:2008,cross/etal:2012} where they are
benchmarked and stored. 

Through the WFAU database, we are able to retrieve any of the 3 levels of
data-product described above: exposures, paw-prints, and/or tiles. We opt to
work with individual paw-print level data. This is primarily because the tile
level data are frequently made up of paw-prints with a range of different
point-spread functions (PSFs), and this can lead to complications later in our
analysis (specifically regarding flux estimation; see Sect.~\ref{sec: photo}).
Therefore, we begin our combination of KiDS and VIKING by first downloading all 
the available stacked paw-prints from the WFAU. We then perform a recalibration
of the individual paw-prints following the methodology of 
\cite{driver/etal:2016} to correct the images for atmospheric extinction
($\tau$) given the observation airmass ($\sec\chi$), remove the exposure-time
($t$, in seconds) from the image units, and convert the images from various Vega
zero-points ($Z_{\rm v}$) to a standard AB zero-point of $30$ \citep[using the
documented Vega to AB correction factors, $X_{\rm AB}$;][]{gonzalez-fernandez/etal:2018} 
which roughly translates to an image gain of ${\rm ADU/{e^-}} = 1$. The
recalibration factor used is multiplicative, applied to all pixels in each
detector image $I$:
\begin{equation}\label{eqn: factor}
I_{\rm new} = I_{\rm old} \times \mathcal{F_{\rm r}}
\end{equation}
and is calculated as: 
\begin{equation}
\begin{split}\label{eqn: recalib}
\log_{10}(\mathcal{F_{\rm r}}) = -0.4&\left[Z_{\rm v}-2.5\log_{10}(1/t)\right.\\
      &\hspace{42pt}-\left.\tau\left(\sec\chi-1\right)+X_{\rm AB}-30\right].  
\end{split}
\end{equation}
This preprocessing of each VISTA detector 
also involves performing an additional background subtraction, which is done using the {\sc SWarp}
software \citep{bertin:2010} with a $256\stimes256$ pixel mesh size and
$3\stimes3$ mesh filter for the bicubic spline. This allows the removal of
small-scale variations in the NIR background with minimal impact on the source
fluxes \citep{driver/etal:2016}. Unlike GAMA, however, we do not recombine the
individual paw-prints into tiles or large mosaics; we choose instead to work
exclusively with the individual recalibrated detectors throughout our analysis. 

After this processing, we perform a number of quality control tests to ensure
that the imaging is sufficiently high quality for our flux analysis. In
particular, we check distributions of background, seeing, recalibration factor
(Equation \ref{eqn: recalib}), and number counts for anomalies. After these
checks, we determined that a straight cut on the recalibration factor was
sufficient to exclude outlier detectors, and thus implement the same rejection
of detectors as in \citet{driver/etal:2016}; namely accepting only 
detectors with $\mathcal{F_{\rm r}}\le 5.0$. 

After this processing and quality control, we transfer the accepted imaging over
to our flux measurement pipeline. Our final sample consists of $301,824$ 
individual detectors across the 5 VIKING filters, drawn from the WFAU proprietary 
database v21.3, which are spread throughout the KiDS-450 footprint. This database, 
however, does not yet contain the full VIKING dataset, as reduction and ingestion 
of the final VIKING data (taken as recently as February 2018) is ongoing. As such, 
the final overlap between the KiDS footprint and VIKING is likely to continue 
to grow with future KiDS+VIKING data releases. 


\section{Photometry and photometric redshifts}\label{sec: photo}

\subsection{9-band photometry}

Multi-band photometry is extracted from the combined KiDS+VIKING data using the Gaussian
Aperture and PSF \citep[GAaP;][]{kuijken:2008,kuijken/etal:2015} algorithm. The algorithm 
generates PSF-corrected Gaussian-aperture photometry that is particularly well 
suited for colour-measurements which are used for estimating photometric redshift. 
{\bbf The GAaP code differs to other standard photometric codes in that it does 
not require images to be pixel nor PSF matched in order to extract matched aperture 
fluxes \citep[such as Source Extractor; ][]{bertin/arnouts:1996}, and it limits flux 
estimation to the typically brighter/redder interior parts of galaxies, unlike 
codes designed for total flux photometry \citep[such as {\sc lambdar}; ][]{wright/etal:2016}. 
GAaP also utilises purely gaussian photometric apertures and PSFs (hence the name), 
and therefore performs the required image gaussianisation prior to flux measurement. }

The algorithm requires
input source positions and aperture parameters, which we define by running
Source Extractor \citep{bertin/arnouts:1996} over our {\sc theli} $r$-band imaging in a so-called 
hot-mode. This refers specifically to the use of a low deblend threshold, which allows 
better deblending of small sources. This choice can have the adverse effect, however, of 
shredding large (often flocculant) galaxies. We choose this mode of extraction as we are 
primarily interested in sources in the redshift range $0.1\lesssim z \lesssim 1.2$, which 
are typically small and have smooth surface-brightness profiles.  Once we have
our extracted aperture parameters, the algorithm then performs a
gaussianisation of each measurement image.  This removes systematic variation
of the PSF over the image and allows for a more consistent estimate of source
flux across the detector-plane. This gaussianisation is performed by
characterising the PSF over the input image using shapelets
\citep{refregier:2003}, and then fitting a smoothly varying spline to the
shapelet distribution. For this reason, it is optimal to provide input images
that do not have discrete changes in the shape of the PSF which cannot be
captured by this smoothly varying distribution. The smooth function is then
used to generate a kernel that, when convolved with the input image, normalises
the PSF over the entire input image to a single Gaussian shape with arbitrary
standard deviation.  

Due to the requirement that the input imaging not have discrete changes in the
PSF parameters, we require that the GAaP algorithm be run independently on
subsets of the data that were taken roughly co-temporally. In the optical this
is trivial; the $1.2$ square degree stacks of jittered observations, called
`pointings', are always comprised of individual exposures with small offsets
that were taken essentially cotemporally, due to the design of the detector
array and survey observation strategy. This, combined with the stability of the
PSF pattern across the field of view that is inherent to observations made at
the Cassegrain focus of a Ritchey-Chr\'etien telescope, means that the stacked
pointings are optimised for use in GAaP. Using the KiDS pointings for optical
flux measurements with GAaP results in at most four flux estimates for any one
KiDS source in the limited corner-overlap regions between adjacent pointings,
or two flux estimates at the pointing edges. However, as in KiDS-450, we mask
these overlap regions such that the final dataset contains only 1 measurement
of all sources within the footprint, rather than performing a combination of
these individual flux estimates. As such, our final flux and uncertainty
estimates in the optical are simply those output directly by GAaP. 

Conversely, the VISTA tiles are particularly sub-optimal for use in GAaP, due
to the large dithers between successive paw-prints which are necessary to fill
a contiguous area on-sky. Stacking such exposures with large dithering offsets
can lead to significant discrete changes in the PSF of the stacked image, and
this problem is exacerbated by the strong PSF variations over the focal plane
inherent to observations made with such a fast telescope. Therefore, in order
to streamline the data handling and avoid non-contiguous PSF patterns we
decided to extract the VISTA NIR photometry from single VISTA detector images
of individual paw-prints, as recommended in \cite{gonzalez-fernandez/etal:2018}. 
In practice,
the paw-print level data are provided as individual detector stacks, rather
than as a mosaic of the telescope footprint. 

\begin{figure*}
\centering
\includegraphics[width=\textwidth]{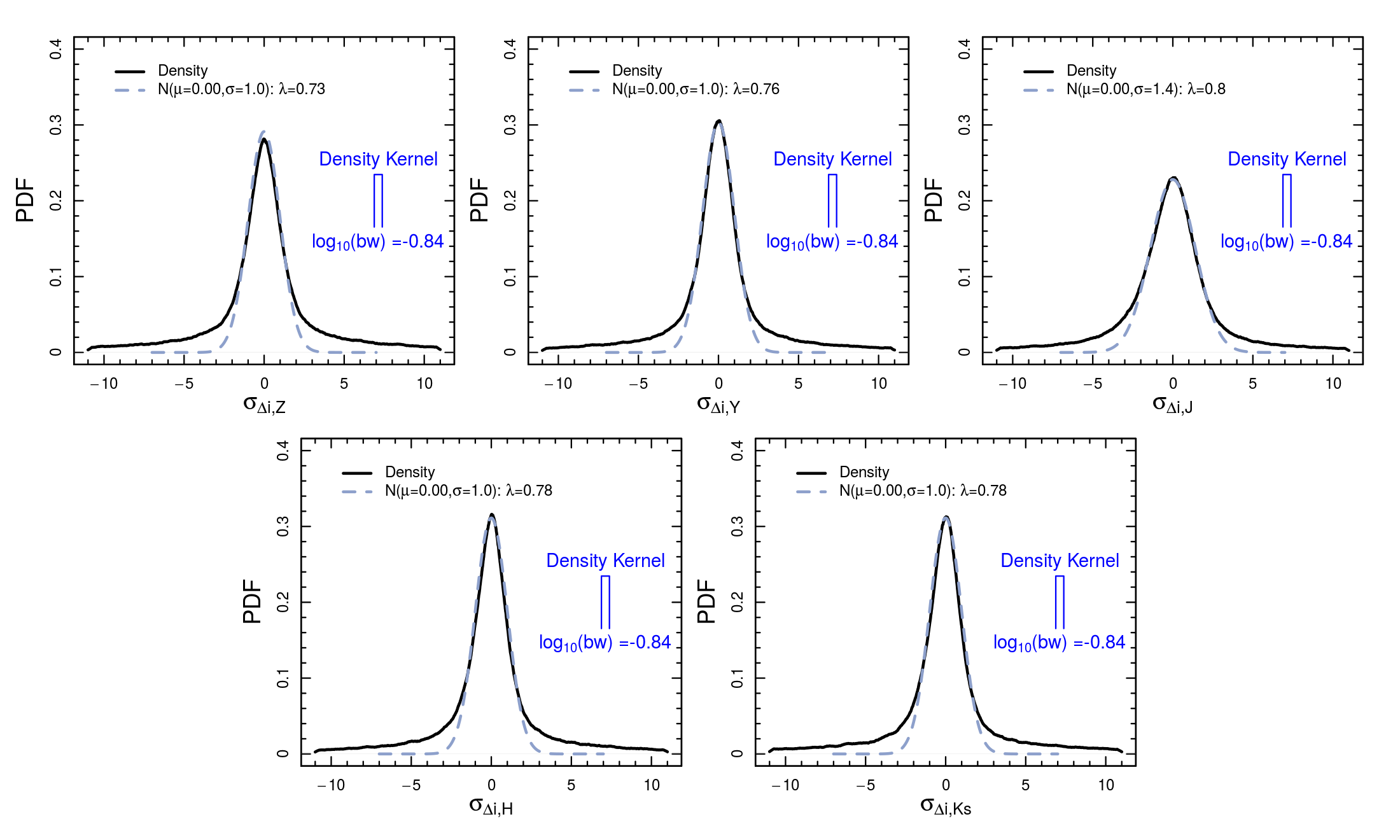}
\caption{The distributions of individual flux measurements with respect to the
final flux estimate and uncertainty in KV450. Here we show per-band PDFs of
$\sigma_{\Delta{i}}$, which demonstrates the accuracy of the final flux
uncertainties for sources in KV450 (see text for details). We overlay on each
distribution a Gaussian model that describes well the core of each
distribution, providing the mean ($\mu$), standard deviation ($\sigma$), and
mixture fraction of the Gaussian given the total PDF ($\lambda$). {\bbf We find
that the final fluxes and uncertainties are generally a good description of the
individual data, with typically $>70\%$ of all individual (per-detector) flux
estimates being well described by the simple gaussian statistics. The wings of
these distributions are caused by the existence of non-gaussian noise
components not encoded by the GAaP uncertainties (\eg\ zero point
uncertainties). We note that} the $J$-band, however, has uncertainties that are
underestimated by roughly $30\%$.  Each panel is annotated with the kernel used
in the PDF estimation, showing the width of the kernel and its log-bandwidth
(bw). 
}\label{fig: internal}
\end{figure*}

Accordingly, we gaussianise the PSF of each paw-print detector in the VIKING
survey separately, and run GAaP on these units.  As there is no one-to-one
mapping between KiDS pointings and VIKING paw-prints, we are required to
associate individual VIKING detectors with overlapping KiDS pointings
on-the-fly. Furthermore, the VISTA dither pattern results in anywhere between 1
and 6 independent observations of a given source within the tile. This
typically results in multiple flux measurements per source and band as most sky
positions within the tile are covered by at least 2 paw-prints in the
$ZYHK_{\rm s}$-bands and at least 4 paw-prints in the $J$-band. Therefore, for
each source we calculate a final flux estimate, $f_{\rm f}$, that is the weighted average of
the $n$ individual flux measurements, $f_{\rm i}$: 
\begin{equation}\label{eqn: finalflux} 
f_{\rm f} = \frac{\sum_{i=1}^n{f_{i}w_{i}}}{\sum_{i=1}^n{w_{i}}},
\end{equation}
where the weight for each source is the individual GAaP measurement inverse
variance $w_{i} = \sigma_{f_{i}}^{-2}$. The final flux uncertainty is
the uncertainty on this weighted mean flux:
\begin{equation}\label{eqn: finalerr} 
\sigma_{f_{\rm f}} = \left[\sum_{i=1}^n \sigma_{f_{i}}^{-2}\right]^{-\frac{1}{2}}.
\end{equation}
To test whether the GAaP flux uncertainties are suitable for use in estimating
the final flux this way, we examine the distribution of sigma deviations
between the final (weighted mean) flux and the individual estimates:  
\begin{equation}\label{eqn: sigmaf}
\sigma_{\Delta{i}} = \frac{ f_{\rm f} - f_{i} }{ \sqrt{n}\,\sigma_{f_{\rm f}} }, 
\end{equation}
where $n$ is the number of flux measurements that went into the computation of
$f_{\rm f}$ and $\sigma_{f_{\rm f}}$. In the limit where the individual flux
uncertainties $\sigma_{f_{i}}$ are perfectly representative of the scatter
between the individual measurements, the distribution of $\sigma_{\Delta{\rm
i}}$ values should be a Gaussian with 0-mean and a standard deviation of 1.
When the flux uncertainties are not representative of the scatter in the
individual measurements, the distribution may deviate in mean, standard
deviation, or both.  In particular, systematic bias in the flux uncertainties
as a function of flux will shift the mean of the distribution away from 0
(and/or give the distribution an obvious skewness), while over- or
under-estimation of the uncertainties as a whole will cause the distribution
standard deviation to decrease or increase, respectively.  Figure \ref{fig:
internal} shows the distributions of $\sigma_{\Delta{i}}$ for each of the
five VIKING bands. The figure shows that our flux uncertainties in the
$ZYHK_{\rm s}$-bands are appropriate and (for the vast majority of estimates)
Gaussian; {\bbf roughly 20\% of our individual flux estimates have a scatter that is
not well described by the simple final Gaussian uncertainty on our flux
estimate, however this is not surprising given that the individual GAaP flux
estimates are purely shot noise; they do not capture the full uncertainty in
cases where there is considerable zero-point uncertainty, sky background,
correlated noise, or other systematic effects which contribute to the flux
uncertainty.} Figure \ref{fig: internal} also demonstrates that our flux
uncertainties tend to be under-estimated in the $J$-band by roughly $30\%$.
Encouragingly, however, the distributions show no sign of systematic bias in
the flux uncertainties, which would be indicated by a significant skewness of
these distributions. 

\begin{figure*}
\includegraphics[width=\textwidth]{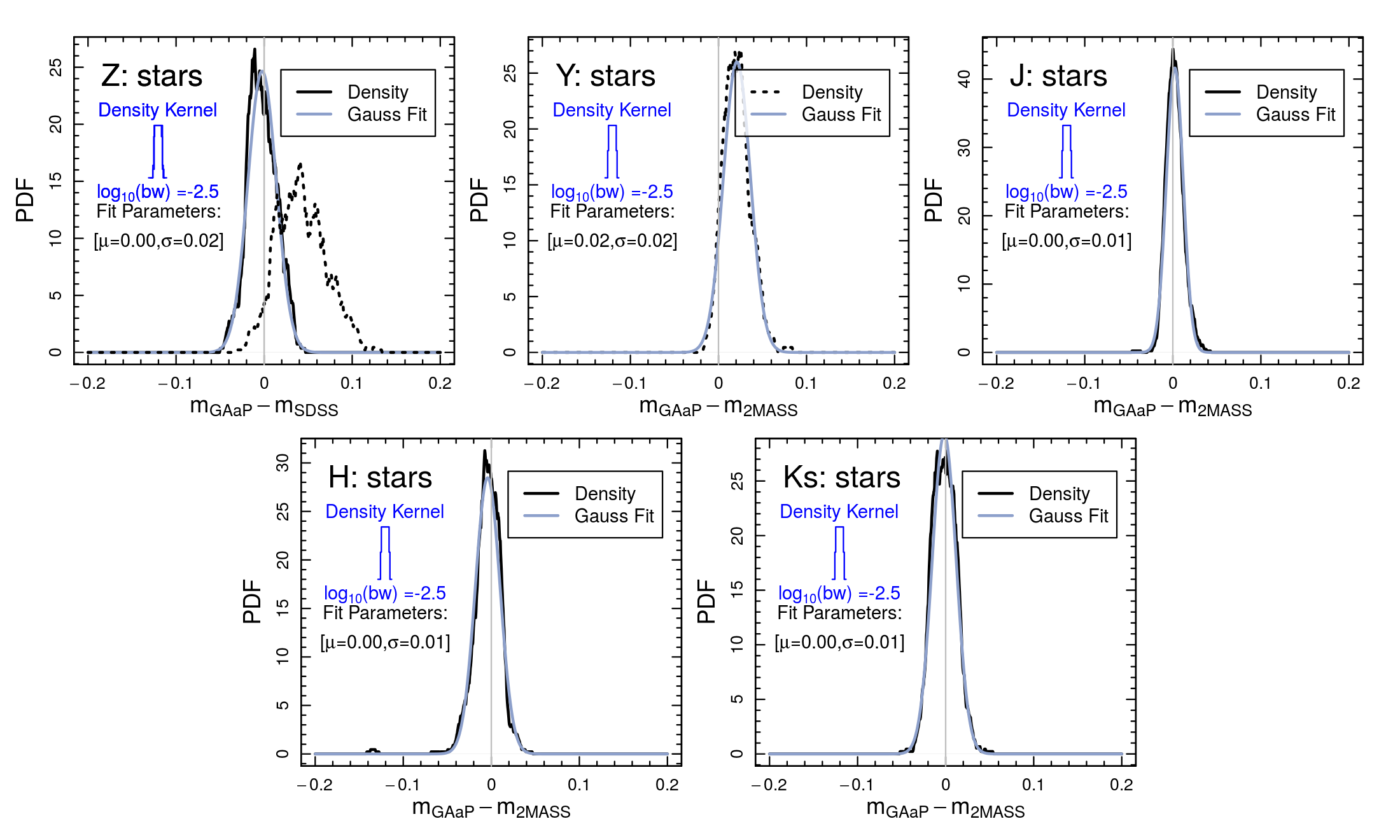}
\caption{\label{fig: KV_vs_2MASS}Photometric comparison of KV450 stellar
photometry in each of the $\ztok$-bands to photometry from SDSS (for
our $Z$-band only, shown as a solid line), and 2MASS in the $\ztok$-bands. 
Note that as 2MASS does not cover the $ZY$-bands, comparisons there
are made using an extrapolation based on the 2MASS $J$-$H$ colour, as described in 
\protect \cite{gonzalez-fernandez/etal:2018}; these are shown here as dashed lines in the $ZY$
comparison panels. We simultaneously fit these distributions with a single
component Gaussian (blue), with the optimised fit parameters annotated. With the 
exception of the Y-band extrapolation (which has a $0.02$ mag residual), 
all directly comparable fluxes are in perfect agreement. } 
\end{figure*}

To verify the calibration of our imaging and flux estimates, we compare our
estimates for a sample of KV450 stars to those measured by SDSS and/or 2MASS.
Stars are particularly useful for this purpose as GAaP yields not only reliable
colours but also total magnitudes for these sources \citep{kuijken/etal:2015},
and therefore we need not be concerned with aperture effects in the flux
comparisons. As the CASU pre-reduction assigns a photometric zeropoint to each
VISTA paw-print based on a calibration with 2MASS, residuals in our multi-band
photometry with 2MASS (particularly in the $JHK_{\rm s}$-bands) would indicate
problems with our pipeline. Similar offsets with respect to SDSS in the
$Z$-band would also be cause for concern. Hence these comparisons are used as
quality control tests, typically on the level of a KiDS pointing. The
distributions of the pointing-by-pointing offsets between our GAaP photometry
and SDSS/2MASS are shown in Fig.~\ref{fig: KV_vs_2MASS}, per band.  The figure
shows the PDFs of these residuals, as well as Gaussian fits to the
distributions. In the $Z$-band, we have two lines: the solid line is a direct
comparison to SDSS, while the dashed line is an extrapolation of 2MASS $J-H$
colours to the $Z$-band. A similar extrapolation is shown in the $Y$-band. Both
of these extrapolations have significant colour-corrections, and so should be
taken somewhat cautiously. Encouragingly, however, in all the cases where we
have fluxes that can be directly compared to one-another (i.e. in all but the
$Y$-band), the direct comparison residuals are centred precisely on 0.
Furthermore, in all cases the fluctuations between pointings are all  within
$\left| \Delta m \right| < 0.02$. 

As a final test of the fidelity of our fluxes, we compare colours of KV450
stars with the same measured in 2MASS, to demonstrate that our observed colours
are consistent with, but less noisy than, those from 2MASS. The distributions
of KV450 and 2MASS $J-H$ and $H-K_{\rm s}$ colours can be seen in
Fig.~\ref{fig: colours}. As expected, the KV450 colours show considerably less
scatter, suggesting that they are a better representation of the underlying,
intrinsic stellar colour distribution \citep{wright/etal:2016}, and are
therefore superior to the colours of 2MASS.  
\begin{figure}
\centering
\includegraphics[width=\columnwidth]{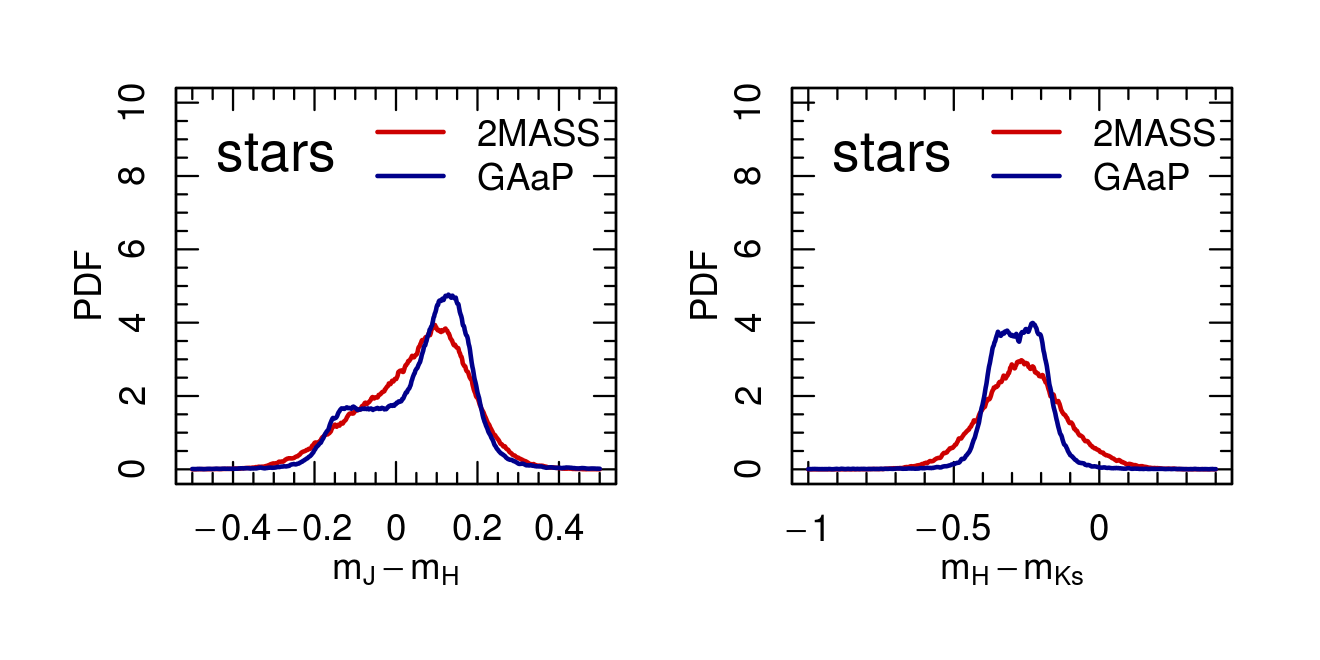}
\caption{Comparison between the colours of KV450 stars and the same sources
measured by 2MASS. The reduction in scatter of the distribution indicates that
the KV450 NIR data have significantly reduced uncertainties. }\label{fig:
colours}
\end{figure}

\begin{figure}
  \centering
  \includegraphics[width=\columnwidth]{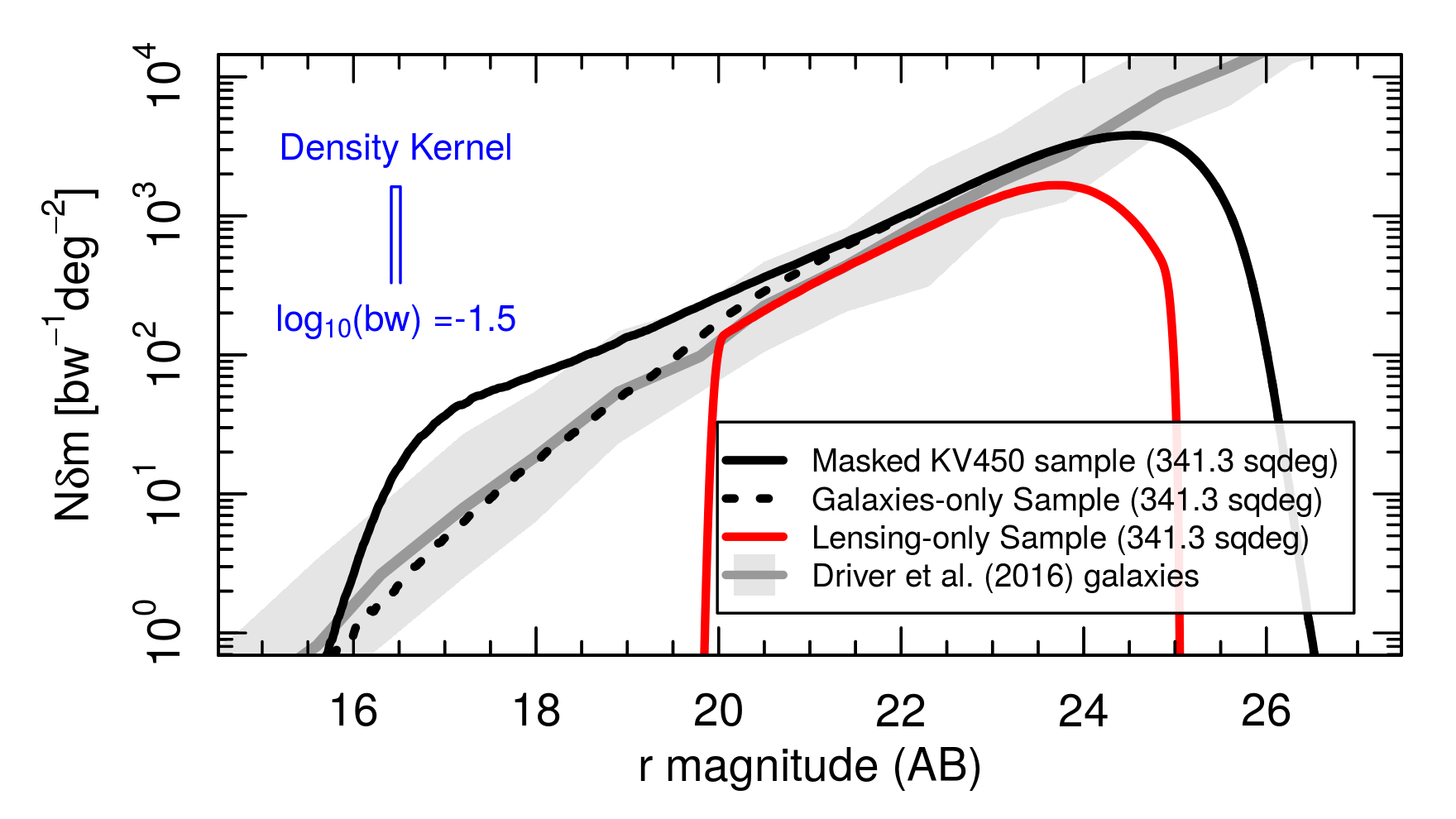}
  \caption{$r$-band number counts for sources in KV450 before ({\em solid
  black}) and after ({\em dotted black}) removal of stars, and for the lensing
  sample ({\em red}). Each of these datasets is presented as raw number density; i.e. 
  the number counts divided by the area of the sample (indicated in the legend), 
  without any additional weighting. We compare these to the galaxy number counts from the
  literature compendium presented in \protect \cite{driver/etal:2016}. The grey region 
  shows the scatter in the data from their literature compendium, while the solid grey 
  line traces the median of their compendium. Our number counts are in good
  agreement with the literature. {\bbf At the bright end our hot-mode
  source extraction leads to a dearth of the brightest galaxies (causing the dashed black line 
  to begin to fall downwards at magnitudes brighter than $r\ssim19.5$).}
  }\label{fig: number counts}
\end{figure}

Now confident that our fluxes are appropriate, we can further verify the
appropriateness of our sample definition and effective-area calculations by
comparing our measured galaxy number counts (in our extraction band, $r$) with
previous works from the literature. Figure \ref{fig: number counts} shows the
$r$-band number counts for the KV450 dataset compared to the compendium of
survey number counts presented in \cite{driver/etal:2016}. We show the KV450
dataset both with and without the removal of stellar sources described in
Sect.~\ref{sec: stars}.  Furthermore, we show the number counts for the sample
of lensing sources used for cosmological inference
\citep{hildebrandt/etal:2018}. The lensing subset is constructed of sources
which are suitable for shape measurement as described in detail in
\cite{hildebrandt/etal:2017}. This lensing sample consists of $13.1$ million
sources, all of which fall within the $r$-band magnitude range $20\lesssim m_r
\lesssim 25$, are unblended, and are resolved.

We see that the all-galaxy sample is lacking in number counts at the brightest
magnitudes; we attribute this to our hot-mode source extraction biasing against
the extraction of the largest, brightest galaxies, as has been noted previously
in earlier KiDS datasets \citep[see, e.g.,][]{tortora/etal:2018}.  Otherwise, the observed
counts of both the all-galaxy- and lensing-only-samples are in excellent
agreement with the literature compendium of $r$-band counts from
\cite{driver/etal:2016}, suggesting that our sample definitions and area
calculations are appropriate. 

Unlike KiDS-450, we also require the final lensing sample to have full 9-band
photometric coverage; i.e. successful photometric measurements are required for every source in all 
9 bands. Table \ref{tab: coverage} provides the photometric measurement
statistics for the lensing sample in KV450, as a function of individual band
and for combinations of bands. The statistics shown are the fraction of sources
with successful GAaP measurements ($f_{\rm good}$) in all 9 bands, for all sources 
that fall both within the area of mutual KiDS+VIKING coverage. 

The table demonstrates that GAaP returns a successful flux measurement for
greater than 98\% of all lensing sources in all bands.  However, as the
failures are different in each band, the full sample ends up with successful
estimates in all 9-bands for greater than 96\% of would-be lensing sources.
Therefore, the requirement of a successful GAaP measurement acts to trim our
final lensing sample down by less than $4\%$. Furthermore, the lensing sample
has significant detections in all 9-bands for over 66\% of the full sample, and
for over 69\% of sources that have no GAaP failures (i.e. where there is data
in all 9-bands). We note, however, that the $u$-band has the lowest number of
significant detections within the dataset, by a considerable margin, but that
{\bbf this is primarily a reflection of physics rather than of the imaging
depth. The rapidly declining nature of galaxy SEDs in this wavelength range at
all redshifts, conspiring with the lower sensitivity of the band compared to,
say, the $g$-band, means that the $u$-band experiences significantly more
non-detected sources than any other band over our redshift window.  Explained
differently: the $i$-band, for example, sees no such dearth in detections
despite being shallower than the $u$-band, courtesy of its probing a typically more 
luminous part of the galaxy SED (at $z\lesssim1$ this is primarily because of 
the flux increase associated with the $4000\AA$ break). } 
Removing the $u$-band from our considerations of detection statistics,  
we find that we have significant detections in the $\gtok$-bands for 
$82\%$ of lensing sources in the dataset. This is a vindication of the combined KiDS+VIKING 
survey design, whereby limiting magnitudes were designed specifically with the goal 
of sampling the 9-band SEDs of the $r$-band selected KiDS sample. 

For completeness, we investigate the cause of the GAaP failures in our dataset. 
These typically occur when either there are data missing, or when the algorithm
is unable to compute the measurement aperture given the image PSF. The latter 
can occur when the PSF full-widths at half-maximum (FWHM) of the measurement 
image is considerably larger than the input (detection) aperture
\citep{kuijken/etal:2015}. As such, the input aperture size can be a source of
systematic bias in the GAaP flux measurement procedure, as smaller input
apertures are more likely to hit the aperture-PSF limit in one of our
non-detection bands. We conclude, however, that this is unlikely to introduce
significant biases into our subsequent analyses as less than 1.2\% of sources
per-band are affected by the GAaP measurement failure. Nonetheless, in 
future releases of KiDS+VIKING data, a recursive flux measurement method will 
be invoked, whereby sources that fail in any band due to this effect are 
subsequently re-measured with an artificially expanded GAaP input aperture. 


After applying the requirement of successful (i.e. $f_{\rm good}$) 9-band photometric
estimation, we finish with a final lensing sample of $\ssim 12.6$ million
sources, which are drawn from an effective area of $341.3$ deg$^2$ (see Sect.~\ref{sec:data}). This is a slight 
reduction in the effective area from KiDS-450 ($360.3$ deg$^2$), however this area will 
recover somewhat in future KiDS+VIKING releases, as the final (full) VIKING area is 
processed and released by CASU (see Sect.~\ref{sec:VIKING}).

\begin{table}
\caption{Measurement statistics for the $13.09$ million lensing
sources that remain after all non-photometry KV450 masks have been applied, per band and as successive
bands are added.  The columns detail the fraction of sources that have
successful GAaP measurements or limits (i.e. where GAaP ran successfully;
$f_{\rm good}$), and the fraction of sources that returned a significant GAaP flux
measurement, and not just an upper limit ($f_{\rm meas}$). 
}\label{tab: coverage}
\centering
\begin{tabular}{ccc}
Band(s) & $f_{\rm good}$ & $f_{\rm meas}$  \\
\hline
$u$  & $0.996$ & $0.794$ \\ 
$g$  & $1.000$ & $0.990$ \\ 
$r$  & $1.000$ & $1.000$ \\ 
$i$  & $1.000$ & $0.954$ \\ 
$Z$  & $0.991$ & $0.983$ \\ 
$Y$  & $0.990$ & $0.965$ \\ 
$J$  & $0.999$ & $0.990$ \\ 
$H$  & $0.989$ & $0.933$ \\ 
$Ks$ & $0.992$ & $0.944$ \\ 
\hline
$\utoi$ & $0.996$ & $0.761$ \\ 
$\utoz$ & $0.987$ & $0.751$ \\ 
$\utoy$ & $0.977$ & $0.732$ \\ 
$\utoj$ & $0.976$ & $0.728$ \\ 
$\utoh$ & $0.967$ & $0.691$ \\ 
$\utok$ & $0.963$ & $0.669$ \\ 
\hline
$\gtok$ & $0.967$ & $0.820$ \\ 
\hline
\end{tabular}
\end{table}

\subsection{Photometric redshifts}\label{sec: photoz}
Photometric redshifts are estimated from the 9-band photometry using the public
Bayesian Photometric Redshift \citep[{\sc bpz};][]{benitez:2000} code. We
use the re-calibrated template set of \citet{capak:2004} in combination with
the Bayesian redshift prior from \cite{raichoor/etal:2014}; hereafter R14. We utilise the
maximum amount of photometric information per source, providing BPZ with both
flux estimates and limits (where available). Finally, input fluxes are
extinction corrected before use within the BPZ code, using \cite{schlegel/etal:1998} dust
maps and per-band absorption coefficients.

We test the accuracy of our KiDS+VIKING photo-$z$ estimates using a large sample
of spectroscopic redshifts collected from a number of different surveys: 
\begin{itemize}
\item zCOSMOS \citep{lilly/etal:2009};
\item DEEP2 Redshift Survey \citep{newman/etal:2013};
\item VIMOS VLT Deep Survey \citep{lefevre/etal:2013};
\item GAMA-G15Deep \citep{kafle/etal:2018};
\item ESO-GOODS \citep{popesso/etal:2009,balestra/etal:2010,vanzella/etal:2008}. 
\end{itemize}
This combined spectroscopic calibration sample, matched to KV450, includes
$>33,000$ sources extending over a 95\% $r$-band magnitude quantile range of
$r\in\left[19.76,24.75\right]$.  Within the sample, $96\%$ of sources have full
9-band photometric information returned by GAaP, and $77\%$ have significant
detections in all 9-bands. This sample is therefore a reasonable match to the
full KV450 dataset, which extends slightly deeper $(r\in\left[20.82,25.18\right])$ 
and has $96\%$ and $67\%$ coverage and detection
fractions, respectively (see Table \ref{tab: coverage}). Detailed information on
the collation of this spectroscopic calibration sample can be found in
\cite{hildebrandt/etal:2018}. 

\begin{figure*}
\includegraphics[width=\textwidth]{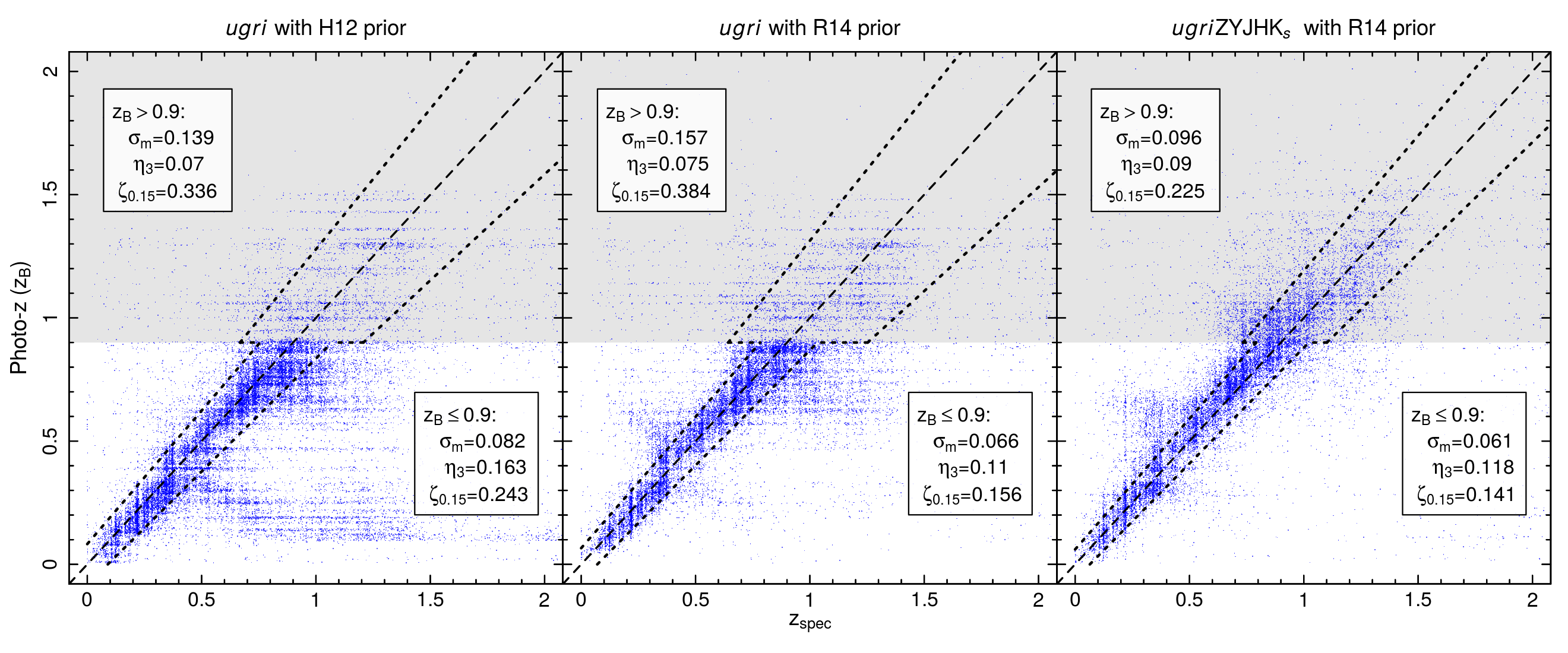}
\caption{\label{fig: zz}Photometric redshifts ($z_{\rm B}$) vs. spectroscopic
redshifts ($z_{\rm spec}$) in the deep calibration fields. \emph{Left:} The
original KiDS-450 photo-$z$ based on $\utoi$-band photometry. \emph{Middle:}
Improved $\utoi$-band photo-$z$ based on the Bayesian prior by
\protect \citet{raichoor/etal:2014}. \emph{Right:} KV450 photo-$z$ based on
$\utok$ photometry as well as the improved prior. The grey region of the figures 
indicate sources beyond the $z_{\rm B}$ limit imposed in the KiDS-450
analysis. Annotated in each panel is: the normalised median-absolute-deviation
($\sigma_{\rm m}$) of the quantity $(z_{\rm B}-z_{\rm spec})/(1+z_{\rm spec})
\equiv \Delta z/(1+z)$, the fraction of sources with $|\Delta
z/(1+z)|>3\sigma_{\rm m}$ ($\eta_3$), and the fraction of sources with $|\Delta
z/(1+z)|>0.15$ ($\zeta_{0.15}$). Each of these quantities is calculated
individually for the sources above and below $z_{\rm B}=0.9$. The value of
$\sigma_{\rm m}$ is also displayed graphically in each panel using the black
dotted lines. Note the significant improvement in all quantities that is seen
when moving from the 4- to 9-band photometry, and in particular that we are now
able to constrain $z_{\rm B} > 0.9$ sources to almost the same accuracy as
those $z_{\rm B} < 0.9$ in the original KiDS-450 dataset.}
\end{figure*}

{\bbf We note here that, importantly, our testing and quality verification of the photo-$z$ 
extend only to the maximum likelihood point-estimate values returned from the redshift fitting 
code: the $\rm Z\_B$ values. This is because for the analyses performed with the photo-$z$ within KiDS, 
only the point-estimates are ever used; the full photo-$z$ PDFs are never considered. Therefore, we note 
here for clarity that the statistics presented here all extend to the Z\_B values only, and no 
quality testing of the full photo-$z$ PDFs is presented. }

Figure~\ref{fig: zz} shows a comparison of our photo-$z$ with the spectroscopic
calibration sample. The figure shows the standard photo-$z$ vs. spec-$z$
distributions for 3 separate photo-$z$ realisations, as well as annotated
statistics for each distribution {\bbf as a function of photo-$z$}. 
These statistics are calculated using the
distribution of $(z_{\rm B}-z_{\rm spec})/(1+z_{\rm spec}) \equiv \Delta
z/(1+z)$ values, and are: 
\begin{itemize}
\item $\sigma_{\rm m}$: the normalised median-absolute-deviation of $ \Delta z/(1+z)$; 
\item $\eta_3$: the fraction of sources with $|\Delta z/(1+z)|>3\sigma_{\rm m}$; and 
\item $\zeta_{0.15}$: the fraction of sources with $|\Delta z/(1+z)|>0.15$. 
\end{itemize}
The three photo-$z$
realisations include the initial KiDS-450 4-band photo-$z$ as presented in
\cite{hildebrandt/etal:2017}, an updated version of the 4-band photo-$z$ using
the R14 prior, and KiDS+VIKING 9-band photo-$z$ (also with the R14 prior).
Comparing the two 4-band photo-$z$ setups, we see that the R14
prior is effective in suppressing outliers in the low photo-$z$ portion of the
distribution by over $30\%$, but shows worse performance at the highest redshifts, 
where the outlier rate and scatter increase by factors of $1.14$ and $1.13$ respectively. The 9-band
photo-$z$, however, shows significant improvement over both 4-band setups. In
particular, the inclusion of the NIR data allows us to constrain photo-$z$ in
the $z_{\rm B}>0.9$ range ($\sigma_{\rm m} = 0.096$) to almost the same level of 
precision as for the $z_{\rm B}<0.9$ sample ($\sigma_{\rm m} = 0.061$), an 
extremely powerful addition to the dataset,
particularly for studies of cosmic-shear where these data carry a very strong
signal. We note that the value of $\eta_3$ increases slightly for the high-$z$ 
portion of the 9-band dataset, however this is primarily because the value of 
$\sigma_{\rm m}$ here is reduced by nearly a factor of two; the higher $\sigma_{\rm m}$ 
in the 4-band cases conceals the non-gaussianity of the distributions, artificially 
reducing the value of $\eta_3$ there. 

We can further motivate the importance of having NIR data for computation of
photo-$z$ by exploring how the statistics which describe the photo-$z$ vs
spec-$z$ distribution vary under the addition of NIR data, as a function of
{\bbf spec-$z$. We note though, that these statistics as a function of spec-$z$ {\em cannot} 
be used for the quantification of photo-$z$ performance for sources selected in discrete 
bins of photometric redshift (such as tomographic cosmic shear bins). Rather, these 
can be used exclusively to demonstrate the influence the additional wavelength information 
has on the data as a function of true redshift. }

\begin{figure}
\centering
\includegraphics[width=\columnwidth]{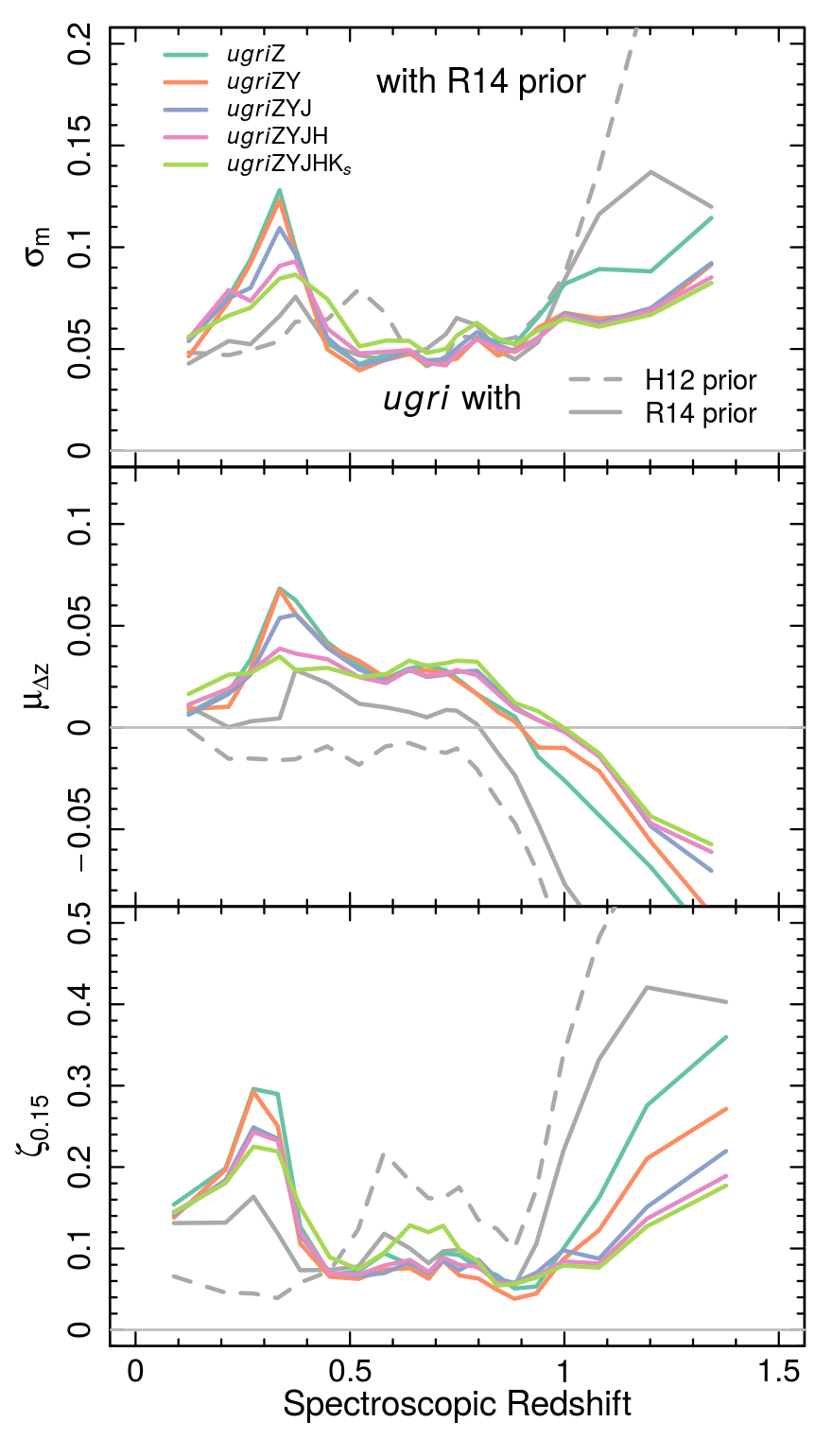}
\caption{Variation in the photo-$z$ vs spec-$z$ distribution parameters as a
  function of {\bbf spec-$z$}, for the 4-band KiDS-450 dataset with two different priors
  ({\em grey lines}), and as a function of NIR photometric information for the
  KV450 dataset ({\em coloured lines}). The three panels show the spread in the
  distribution, determined by a running normalised-median-absolute-deviation from
  the median ($\sigma_m$; {\em top}), the median bias in the photo-$z$
  distribution ($\mu_{\Delta z}$; {\em middle}), and the fraction of sources with
  $|\Delta z| / (1+z_{\rm spec}) > 0.15$ ($\zeta_{0.15}$; {\em bottom}). The
  addition of the \protect \cite{raichoor/etal:2014} prior to the 4-band data
  causes significantly better behaviour at low $z_{\rm B}$, while the addition of NIR
  data improves the population consistency and scatter in particular at high
  $z_{\rm B}$. {\bbf The same properties as a function of photo-$z$ and magnitude are 
  given in Figures \protect \ref{fig: photoz stats 2} and \ref{fig: photoz stats 3}.}
}\label{fig: photoz stats}
\end{figure}

{\bbf Figure~\ref{fig: photoz stats} shows the change in our 3 parameters
of interest as a function of $z_{\rm spec}$, for changes in the prior (for the
4-band KiDS-450 data in grey) and under addition of NIR data (using only
the R14 prior in colours). The 3 parameters in the figure are as follows:
$\sigma_{\rm m}$, the median bias in $\Delta z/(1+z)$ ($\mu_{\Delta z}$), and
$\zeta_{0.15}$. Each parameter is shown using a running median in $20$ equal-N
bins of $z_{\rm spec}$. The equivalent figure with bins constructed as a function of 
$z_{\rm B}$ and $r$-magnitude are given in Figures \ref{fig: photoz stats 2} and 
\ref{fig: photoz stats 3}. 

The statistics as a function of $z_{\rm spec}$ demonstrate that it is the
combination of all 9-bands which performs the best across both the full gambit
of statistics and the redshift baseline. The addition of the $Z$-band causes a
clear improvement, in all statistics, over the $4$-band case when we move
beyond $z_{\rm spec}=0.9$. This is because at $z_{\rm spec}=0.9$ 4000Å-break 
flux enters the i-band, and in the $4$-band case is therefore poorly sampled 
and becomes sensitive to noise fluctuations. With the addition of the $Z$-band, however,
sampling of this flux is more robust and the statistics unilaterally improve. 
There are further improvements with the addition of subsequent bands: the Y-band causes a large
reduction in scatter at z>1, because the same post-4000Å-break flux is now
sampled by 2 or more bands (further decreasing the influence of noise). This benefit 
then saturates (subsequent bands do not improve the high $z_{\rm spec}$ scatter), 
however the story nonetheless continues. The $J$- and $H$-bands are primarily
responsible for a reduction of outliers at $0.2<z_{\rm spec}<0.4$, where a model degeneracy
(which is considerably {\em worse} after the inclusion of the $Z$-band) populates a cloud 
which can be seen in the photoz-specz distribution (Figure \ref{fig: zz}). Finally
the $K_{\rm s}$-band helps bring the low-$z_{\rm spec}$ scatter down further, and also produces the
lowest overall high-$z_{\rm spec}$ outlier rate. Indeed, the outlier rate at $z>0.9$ reduces
continuously with the number of bands added. For these reasons, we conclude that the full complement
of the 9-band data is what is required for the best performance, especially at $z_{\rm spec}>0.9$. 
}

\subsubsection{Binned by $z_{\rm B}$} 
\begin{figure}
\centering
\includegraphics[width=\columnwidth]{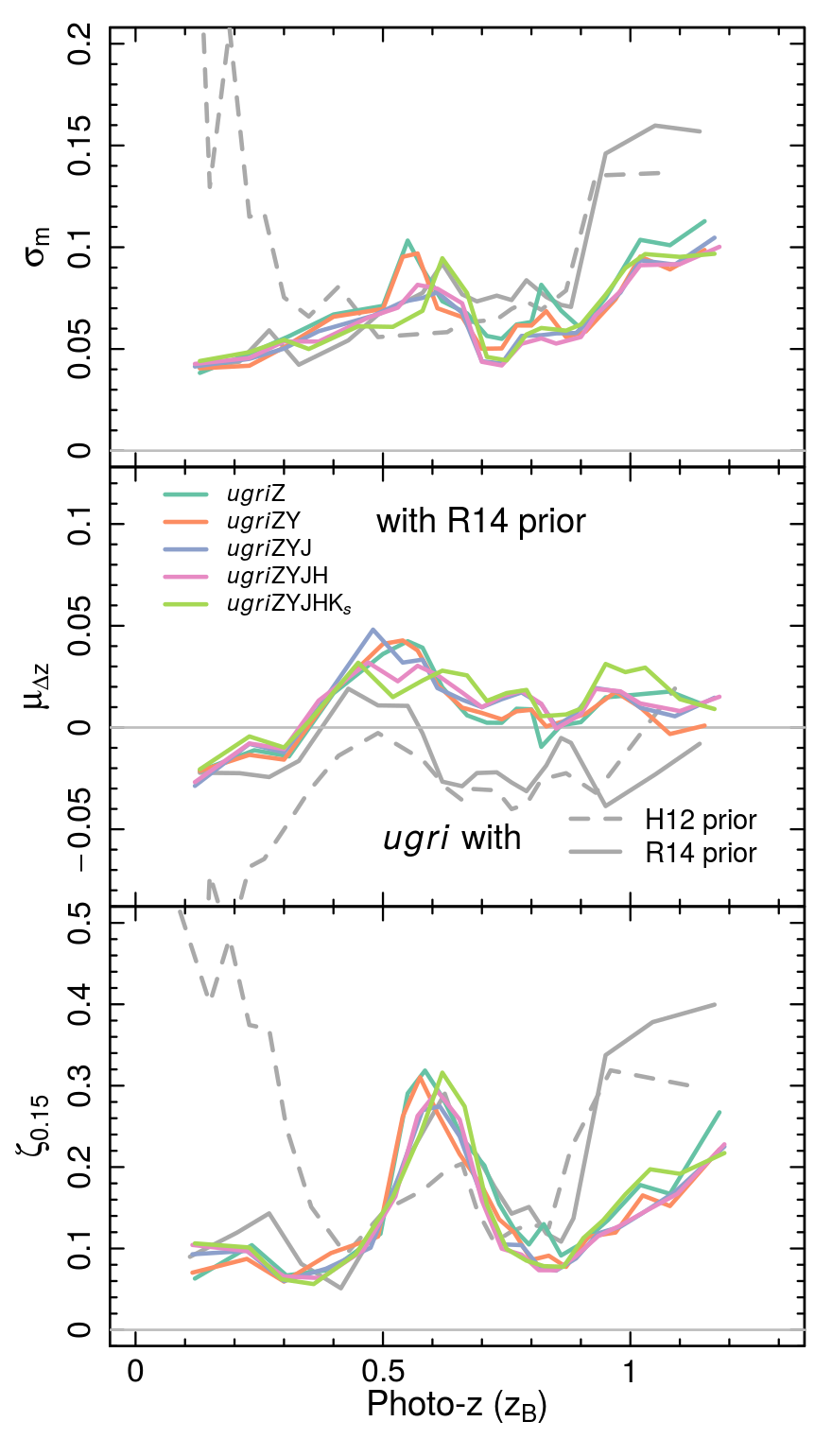}
\caption{{\bbf Variation in the photo-$z$ vs spec-$z$ distribution parameters as a
  function of photo-$z$. The figure is constructed the same as Figure 
  \protect \ref{fig: photoz stats}. }
  }\label{fig: photoz stats 2}
\end{figure}

\begin{figure}
\centering
\includegraphics[width=\columnwidth]{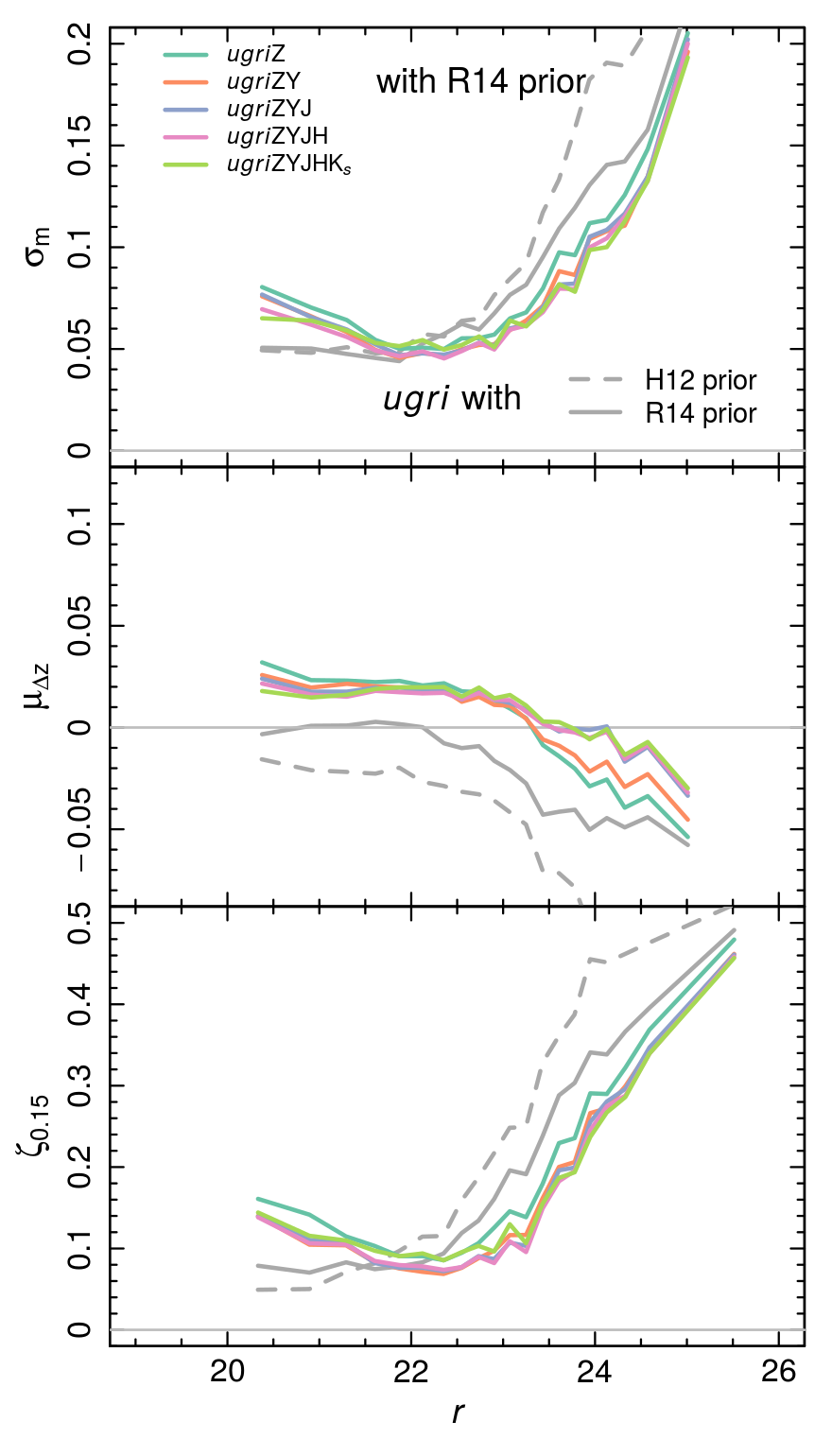}
\caption{{\bbf Variation in the photo-$z$ vs spec-$z$ distribution parameters as a
  function of $r$-magnitude. The figure is constructed the same as Figure 
  \protect \ref{fig: photoz stats}. }
  }\label{fig: photoz stats 3}
\end{figure}

{\bbf
Again, we note that these trends shown above and in Figure \ref{fig: photoz stats} are not directly transferable to a
sample defined as a function of photo-$z$. We show the influence of the individual bands on photometrically defined
samples in Figure  \ref{fig: photoz stats 2}.  

Looking at the effect of the updated prior on the 4-band
photo-$z$ statistics, we see that the new prior has the effect of greatly
reducing scatter at low $z_{\rm B}$, while also reducing bias across essentially
all $z_{\rm B}$. There is also a slight increase in the outlier rate with the new
prior at intermediate and high $z_{\rm B}$, but this is minor compared to the  
significant decrease at $z_{\rm B} < 0.4$.

When combining the NIR data (starting with the $Z$-band) with the 4-band
photometry, we see an immediate improvement in the distribution scatter and
outlier rate at high $z_{\rm B} > 0.7$. In this range, when incorporating all NIR bands, 
we see decreases in scatter of between $30$ and $60$ per cent, over the 
4-band R14-prior case. 
Of particular note is the effect of adding the 
NIR-bands to the outlier rate at $z_{\rm B} > 0.7$. Here the added data reduce the 
observed outlier rate by a factor of $\ssim 2$. Overall, the distributions demonstrate that
NIR data as a whole are extremely useful in constraining photo-$z$ for sources in the
redshift range $0.7 < z_{\rm B} < 0.9$, and are invaluable for the estimation of
photo-$z$ at $z_{\rm B} > 0.9$. 
}

\subsubsection{Binned by $r$-magnitude} 
{\bbf
The introduction of the NIR data actually creates an increase in the observed scatter and outlier rate 
for sources at the brightest magnitudes. However beyond $r=22$, both the scatter and outlier rate reduce 
to levels superior to the $4$-band data. With the addition of subsequent NIR bands (i.e. $\ytok$), we see 
a essentially continual improvement in all statistics over the whole magnitude range. Otherwise, the 
distributions show the expected behaviour of photo-$z$ accuracy as a function of noise; the fainter (and so 
noisier) data exhibit higher scatter in their photo-$z$ and similarly higher outlier rates. We note, though, 
that this definition of outlier rate becomes somewhat nonsensical beyond $r\sim25$, where the scatter 
of the distribution reaches $\sim 0.15$; i.e. the outlier criterion. 

Of particular interest is the reduction in bias that is seen with the introduction of the $J$-band at 
$r \gtrsim 23$. The sources here show the largest bias in the $4$-band case, and this bias is only partially reduced  
in the $Z$ and $Y$ band cases. However the introduction of the $J$-band data causes the bias to reduce 
somewhat. The addition of the $H$- and $K_{\rm s}$-bands do not further reduce the scatter at the faint end, however 
do produce slightly lower biases at the brightest magnitudes. Again, we therefore conclude that the combined 
$9$-band dataset is therefore that which provides the best overall statistics.
}

\subsubsection{Photo-$z$ distributions per field}
\begin{figure}
\centering
\includegraphics[width=\columnwidth]{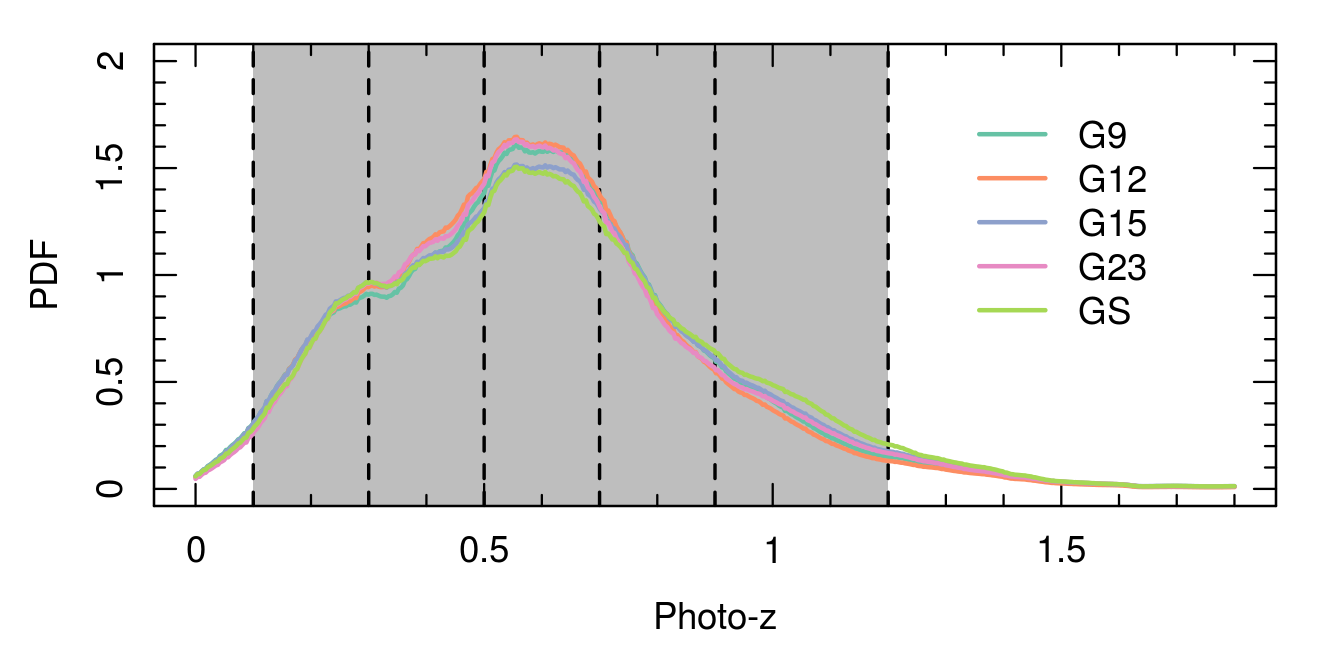}
\caption{{\bbf Distributions of photo-$z$ within each of our 5 survey fields (shown in Figure
\protect\ref{fig:footprint}). The figure shows the PDF estimated using a width$=0.1$ top-hat kernel for each of the 5
fields (coloured lines). The tomographic bins used in KiDS are shown by the grey shading and black dashed lines. 
Sources plotted here are those within the KV450 lensing selection (Figure \protect\ref{fig: number counts}) and
with an additional $r\leq 23.5$ magnitude selection, to remove the effect of variable depth on the comparison.  The
figure demonstrates that, in a like-for-like comparison between the fields, the KV450 photo-$z$ are homogeneous. 
}
}\label{fig: photoz per field}
\end{figure}

{\bbf
Another quality check for the homogeneity of the data is to compare the distributions of photometric redshift in each of
our 5 fields (shown in Figure \ref{fig:footprint}). We compare the
distribution of all photo-$z$ estimates for sources within our lensing sample (Figure \ref{fig: number counts}) with
$r\leq23.5$. These two cuts allow us to compare the photo-$z$ distributions per field for samples of known non-stellar sources 
in a regime agnostic to the effects of variable depth from the comparison; a like-for-like comparison. These
distributions per field are shown in Figure \ref{fig: photoz per field}. We can see from the distribution that the
fields are in very good agreement, with only GS appearing slightly deeper than the other 4 fields. As such, we conclude
that the photo-$z$ among the different KV450 fields demonstrate satisfactory homogeneity. 
}


\section{Higher-order data products} \label{sec: masses} 
We can subsequently utilise our photo-$z$ estimates to
derive higher-order data-products. For this work, we choose to explore the
rest-frame photometric properties of a selection of KV450 sources, as well as
examine the fidelity of integrated properties, namely stellar masses. In order
to explore these properties we perform template-fitting to the broad-band
spectral energy distributions (SEDs) of each KV450 source, while maintaining 
a fixed redshift at the value of $z_{\rm B}$.  

\subsection{SED fitting}\label{sec: SEDs}
To estimate the rest-frame properties of our KV450 sources, we perform SED
fitting with the {\sc Le Phare} \citep{arnouts/etal:1999,ilbert/etal:2006}
template-fitting code, using a standard concordance cosmology of $\Omega_{\rm
m}=0.3$, $\Omega_\Lambda = 0.7$, H$_0$ = $70$ km s$^{-1}$ Mpc$^{-1}$,
\cite{chabrier:2003} IMF, \cite{calzetti/etal:1994} dust-extinction law,
\cite{bruzual/etal:2003} stellar population synthesis (SPS) models, and
exponentially declining star formation histories. Input photometry to {\sc Le
Phare} is as described in Sect.~\ref{sec: photo}, including the per-band
extinction corrections as used in BPZ. We fix the source redshift to be the
value of $z_{\rm B}$ returned from BPZ. We opt to fit SEDs to all $> 45$
million KV450 sources, regardless of masking, so that any/all subsequent
subsamples of KV450 data may incorporate our stellar mass estimates. This
requires that we also allow SEDs to be fit with QSO and stellar templates, for
which we use the internal {\sc Le Phare} defaults. 

\subsection{Star-galaxy separation}\label{sec: stars}
One advantage of fitting all photometric sources in this way is that we are
able to use the higher-order data products to assist with star-galaxy
separation. In particular, by fitting all sources with templates for QSOs,
stars, and galaxies, we are able to identify stellar contaminants that
otherwise would make it into our overall sample. To do this, we identify all
sources that are best fit by a stellar template in {\sc Le Phare} and which have
an angular extent that is point-like; specifically a flux-radius of $0.8$
arcseconds or smaller. Using this simple cut, we are able to produce an
exceptionally clean galaxy-only sample (as shown in Fig.~\ref{fig: number
counts}). We note, however, that this rejection has no effect on the lensing
sample as the high-fidelity point-source rejection that is already performed
during shape-fitting is very effective at removing stellar contaminants. {\bbf
Indeed, all sources that are identified as stars using our SED based selection
are also flagged as stars during shape-fitting.  Furthermore, for the sources
with $g \leq 21$, we can cross-reference our stellar classification with that from
the GAIA DR2 \citep{GaiaDR2} point-source catalogue. 
Comparing to GAIA we find that $99.1\%$ of our sources
classified as stars (and which are brighter than the $g \leq 21$ GAIA magnitude
limit) are also classified as stars by GAIA\footnote{Should the GAIA catalogue have 
contamination by truly-extended galaxies, this would indicate a galaxy contamination within 
our star sample in the same proportion.}. Again, this further increases our
confidence in the stellar classification possible using our SED products. 
}


\subsection{Stellar mass estimates} 
For this work, we are primarily interested in assessing the fidelity of stellar
masses that are estimated using the KV450 dataset. Our stellar masses,
estimated by {\sc Le Phare},  are calculated as the mass of stars required to
produce the observed galaxy SED given the best-fit stellar population, assuming the 
combination of models given in Sect.~\ref{sec: SEDs}. Therefore, in order to
recover a fair estimate of the galaxy stellar mass, the observed SED must be
representative of the total light emitted from the galaxy.
Our aperture fluxes, however, have been intentionally optimised for
high-fidelity colours, rather than for the recovery of total fluxes. This means
that our mass estimates here will be systematically below what would be
recovered with a total flux aperture, primarily as a function of source size.
In order to remedy this systematic effect, we opt to use our quasi-total Source
Extractor AUTO flux estimates (measured during our initial source extraction)
to correct our masses. To do this, we implement a correction akin to the
fluxscale correction discussed in \cite{taylor/etal:2011,wright/etal:2017},
although the implementation there was designed to correct for systematic bias
in \cite{kron:1980} apertures for changing galaxy profile shapes. 

Here our fluxscale factor, ${\mathcal F}$, is a multiplicative correction
defined as the linear ratio of the quasi-total Source Extractor $r$-band AUTO
flux to the non-total GAaP $r$-band flux: ${\mathcal F}= f_{\rm AUTO}/f_{\rm
GAaP}$. This correction is applied {\em post-facto} to the {\sc Le Phare}
stellar mass estimates. The correction devised is such that our final SEDs will
be fixed to the AUTO flux estimate, and our SEDs themselves will be reflective
of the flux contained within the GAaP apertures. This can lead to systematic
biases.  For example, if there are significant colour gradients within the
galaxies in our sample, such that the colours within and beyond our apertures
differ considerably, then our SEDs will tend to be non-representative of the
true integrated galaxy spectrum. Admittedly, however, this is only likely to be
a significant effect for galaxies whose size is significantly larger than the
PSF; i.e. low-redshift galaxies for which our analysis pipeline is already
sub-optimal. 

For validation purposes, we compare our fluxscale-corrected stellar mass
estimates to those also estimated by GAMA \citep{wright/etal:2017} and
G10-COSMOS \citep{andrews/etal:2017,driver/etal:2018} in Fig.~\ref{fig: stellar
masses compar}. Both of these studies utilise spectroscopic redshifts, and implement the same cosmology, SPS
models, dust-law, and IMF as used in this work when estimating stellar masses. They also 
use total matched aperture fluxes. These similarities allow direct comparison
of our mass estimates, despite the use of different algorithms and wavelength
bandpasses for the mass estimation. We perform this comparison both for the
KV450 masses described above and for masses estimated in the same way but
utilising only 4-band photometric information (i.e. the KiDS-450 equivalent
masses). The GAMA dataset here is sky-matched to our KiDS-450/KV450 dataset
within a $1$ arcsec radius, for GAMA galaxies with redshift $z\geq0.004$, GAMA
redshift quality flag nQ$>2$, and for KiDS-450/KV450 sources with $z_{\rm B}<0.7$ (so
as to avoid spurious matches to the much deeper KiDS-450/KV450 catalogues).
The G10-COSMOS sample is subset such that it contains only sources with spectroscopic redshifts
(i.e. those with G10-COSMOS flag $z_{\rm use}\leq3$) and is also sky-matched to
KiDS-450/KV450 with a $1$ arcsecond radius. Note that there is no requirement
for consistency between matched sources photo-$z$ and spec-$z$ values. As such,
the scatter here is a reflection of the scatter in the mass estimates due to,
jointly, systematics in our photometric data and photo-$z$ estimation. 

\begin{figure*}
\centering
\includegraphics[width=\textwidth]{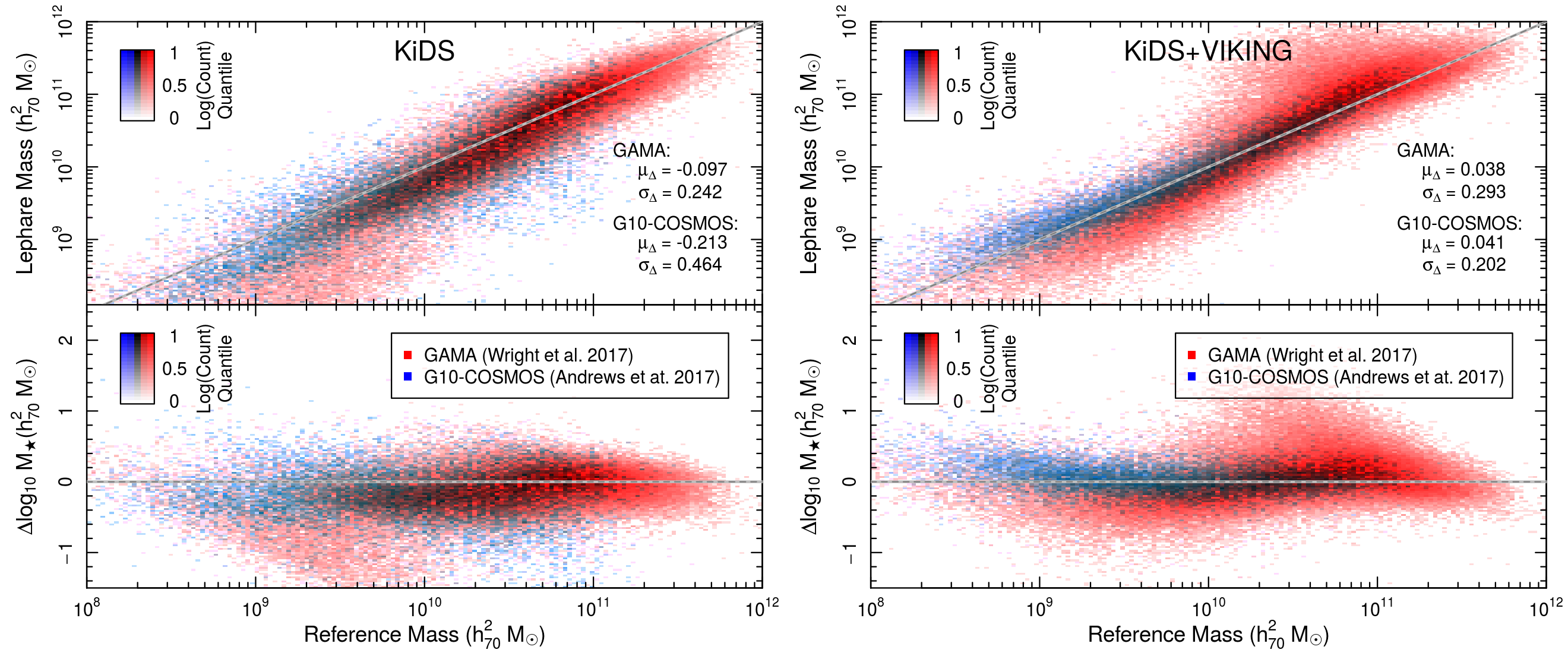}
  \caption{Comparison between stellar mass estimates from both KiDS-450 ({\em left}) and KV450 ({\em right})
  with those from both the GAMA and G10-COSMOS samples, for those sources which
  overlap. KiDS-450 masses here are derived using the KiDS-450
  photo-$z$ {\bbf (\ie with H12 prior)} and only $\utoi$ photometry. Both KiDS datasets are shown with
  masses which have been corrected using our fluxscale parameter. Sources in
  the comparison samples are selected for comparison only if their masses have
  been estimated using spectroscopic redshifts. The figure demonstrates the
  significant improvement in mass estimates that is made when using 9-band
  photometric information. In particular, significant reduction on the scatter
  of the deep G10-COSMOS dataset is particularly important. Scatter in the highest-mass GAMA 
  sources is due to the updated photo-$z$ prior, which is optimised for sources fainter than many 
  high-mass GAMA galaxies.
}\label{fig: stellar masses compar}
\end{figure*}

We see that the KiDS-450 masses show significant scatter in the comparison
distributions (Fig.\ref{fig: stellar masses compar}, left panels), particularly for the COSMOS
dataset which extends to
significantly higher redshift than the GAMA sample ($\sigma = 0.464$). 
Conversely, we see very good 
agreement with the same sample when using masses derived with KV450; 
$\sigma = 0.202$. We note that the scatter in the mass comparison with the GAMA 
sample increases slightly when moving from KiDS-450 to KV450. This increase in scatter between masses 
estimated in KV450 and by GAMA is slightly larger than the typical scatter induced by slightly different mass 
estimation methods \citep[$\ssim0.2$ dex; see][for a detailed discussion of such comparisons and
systematic effects]{wright/etal:2017}, and is induced by the updated photo-$z$ prior implemented here (Sect.~\ref{sec: photoz}). 
This is not surprising, given that this prior is optimised for analysis of the full KiDS sample, which is 
COSMOS-like. The variation between KV450 and GAMA is highly correlated
with systematic differences between the GAMA spec-z and KV450 photo-$z$, which
shows roughly a factor of two stronger bias than we see in the main survey
spectroscopic calibration sample (i.e. Fig.~\ref{fig: zz}), again due largely to our updated prior. 
Importantly, we see no such systematic variations in our comparisons with G10-COSMOS (in mass
or photo-$z$) for KV450. This is in stark contrast to the significant bias and 
scatter that is evident in the KiDS-450 to G10-COSMOS comparisons. 
In particular, we note that the bias in the G10-COSMOS comparison decreases by nearly an 
order of magnitude when moving from KiDS-450 ($\mu_\Delta=-0.213$) to KV450 
($\mu_\Delta=0.041$). Furthermore, we note that the scatter in the comparison 
between KV450 and G10-COSMOS is reduced to $\sigma_\Delta=0.208$; consistent 
with the $0.2$ dex typical uncertainty induced by different mass estimation methods agnostic 
of variations in input photometry and redshifts. As such, we conclude that, for our KV450
sample, the 9-band stellar mass estimates are equivalent in quality to those
that can be estimated using significantly more accurate spectroscopic redshift
surveys. 

\section{Stellar Mass Function}\label{sec: mass function}
Given the accuracy of our observed stellar mass estimates when compared to the
G10-COSMOS survey, we are prompted to explore whether we can reproduce
complex redshift-dependent mass functions using these estimates.  Such mass
functions typically require spectroscopic redshift estimates and/or
high-accuracy photo-$z$ estimates derived from $20+$ broad and narrow
photometric bands
\citep[see, e.g.,][]{andrews/etal:2017,davidzon/etal:2017,wright/etal:2018}.
However, given the apparent fidelity of our mass and photo-$z$ estimates, we
wish to explore whether we can derive sensible mass-evolution distributions
from our relatively low-resolution photo-$z$ estimates alone. 

\begin{figure}
\centering
\includegraphics[width=\columnwidth]{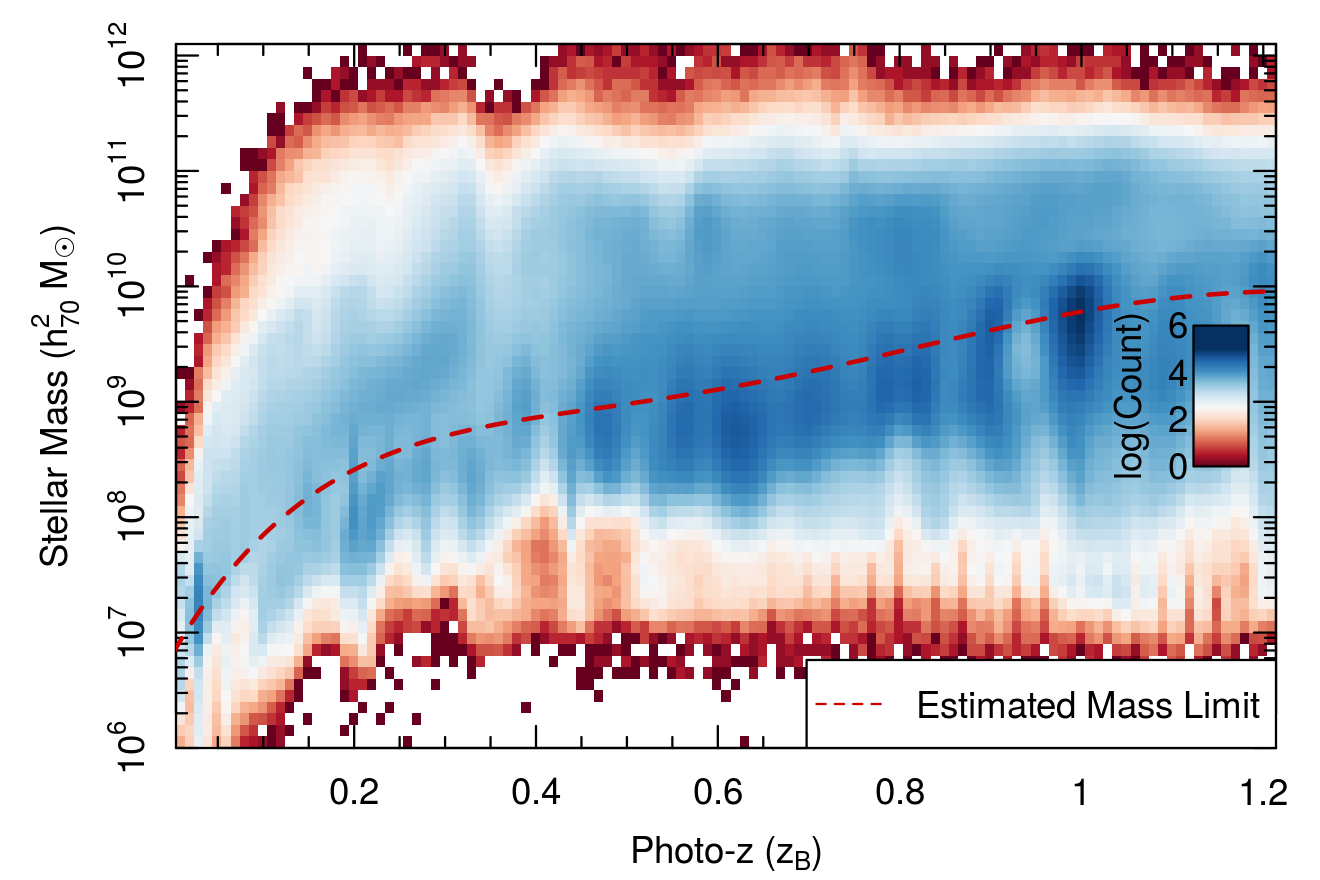}
\caption{Distribution of all KV450 galaxy stellar mass estimates as a function
of photo-$z$. The data is shown as a 2D-histogram with logarithmic scaling.
The distribution is fairly consistent with what is expected of a magnitude
limited galaxy sample, although the incompleteness at low-z is worth noting.
The distribution is fairly uniform above the mass limits (red dashed
line). Below the limits we see signs of systematic incompleteness and redshift 
focussing (caused by the typically noisier data there). }\label{fig: stellar masses}
\end{figure}

Fluxscale-corrected stellar masses from {\sc Le Phare} are shown in
Fig.~\ref{fig: stellar masses} for all galaxies in the KV450 footprint, as a
function of $z_{\rm B}$.  
The distribution shows an underdensity of high-mass sources at
low-redshift, and also a considerable amount of structure as sources approach
the detection limit. This structure is a form of redshift focussing, and is
caused by sources systematically dropping below the detection limit in particular 
bands as a function of galaxy SED shape. Otherwise, the distribution is well bounded 
and fairly uniform, showing little evidence of photo-$z$ dependent biases. 

We wish to use this distribution of stellar masses to estimate a series of 
volume-complete galaxy stellar mass functions (GSMFs) for the KV450 dataset. 
To do this, we first define the mass limit of the dataset as a function of 
photo-$z$. We take the same method of estimating the mass limits as described 
in \cite{wright/etal:2017}, using the turn-over points in both number counts and 
photo-$z$ to estimate the mass-completeness limit. {\bbf Briefly, 
the mass limits as a function of photo-$z$ are constructed assuming that any 
observed down-turn in number-density is due exclusively to incompleteness; \ie\ 
that the mass function, over the redshifts and masses probed here, has no
true down-turn. Using this assumption we estimate the completeness limit as 
a function of photo-$z$ as being the point at which either comoving number 
density and/or stellar mass number density starts to fall.  The procedure is
shown graphically in Figure C1 of \cite{wright/etal:2017}. The calculation of
the completeness limit is} done in a series of overlapping bins of photo-$z$
and stellar mass, and the resulting limit estimates are fit with a fifth-order
polynomial. This derived mass limit is shown in Fig.~\ref{fig: stellar masses}
as a dashed red line. The mass limit can be seen to effectively select against
sources in the redshift-focused low-SNR portions of the distribution, and
suggests that the mass estimates of KV450 can be considered to be volume
complete down to $M_\star\geq 10^{10}M_\odot$ for sources with $z_{\rm
B}\leq1$. 

Using these mass limits, we define a series of volume-complete bins in stellar
mass and redshift, and calculate the resulting mass functions in these bins.
These are shown in Fig.~\ref{fig: massfunc}. For our binning, we choose to use
the same tomographic redshift limits as are implemented in our cosmological
analysis \citep{hildebrandt/etal:2018}, out to $z_{\rm B}=1.2$.  The mass functions are calculated using a
simple volume calculated using the survey area and the redshift limits
annotated in each bin, and we show the mass functions derived with and without
the implementation of the fluxscale correction, for reference.  For comparison,
we also show the model evolutionary mass functions presented in \cite{wright/etal:2018}, 
derived using a compilation of consistently analysed GAMA, G10-COSMOS, and 3D-HST data 
over the redshift range $0.1 \le z \le 5$. For demonstration, the \cite{wright/etal:2018} model 
is shown both as the model expectation at the mean redshift of the bin (grey line), and as the
range of model values (grey shading) that would be expected when allowing for:
photo-$z$ bias $|\Delta z_{\rm B}| \leq 0.2$, additional systematic bias in our
stellar mass estimation ($|\Delta
M_{\star,\rm{sys}}| \leq 0.2$ dex), and Eddington bias ($|\Delta
M_{\star,\rm{edd}}|= 0.2$ dex). 

The first photo-$z$ bin shows a mass function that has a clear deficit in number 
density for the highest mass sources. This deficit, we argue, is again caused by 
our pipelines optimisation for small-angular scale sources: the largest sources on sky 
will also be the most massive at low redshift, and our analysis methods are biased 
against accurate extraction of these sources. In the subsequent bins, however, 
the mass functions from our sample are in good agreement with the evolutionary 
model of \cite{wright/etal:2018}. This is particularly noteworthy, given the coarseness 
of our photo-$z$ estimation and that no correction for the redshift distribution 
bias \citep[such as is done in cosmic shear analyses; see][]{hildebrandt/etal:2017} 
has been attempted. The mass functions, however, clearly 
suffer from considerable Eddington bias in their masses (i.e. our mass functions are 
biased toward higher masses).

\begin{figure*}
  \centering
  \includegraphics[width=\textwidth]{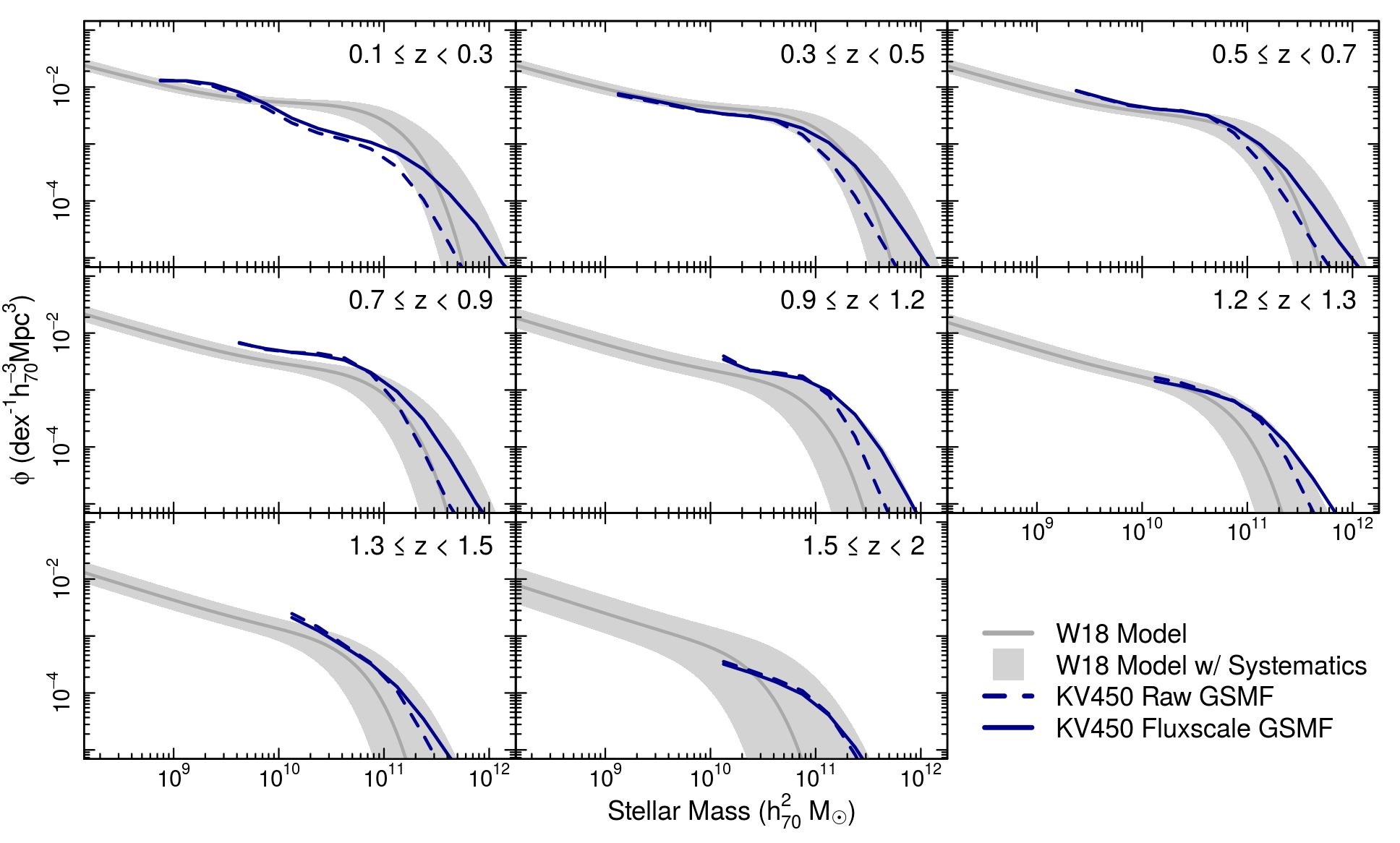}
  \caption{Galaxy stellar mass functions for the KV450 dataset, shown with
  (blue, solid) and without (blue, dashed) the fluxscale factor incorporated,
  compared to the mass function model given in \protect \cite{wright/etal:2018}
  (dark grey). KV450 lines are shown only over the range where we believe the
  mass function to be volume complete. To ensure a fair comparison, we also
  show the model mass function when allowing for uncertainty in photometric
  redshift ($|\Delta z|\leq 0.02$) and Eddington bias ($|\Delta M_\star|\leq
  0.2$) in grey. The KV450 mass function shows a significant deviation from the
  expectation in the lowest redshift bin, which we attribute to a bias against
  selecting the largest-angular-size galaxies in our analysis (see Section
  \ref{sec: mass function}). In all other bins the agreement with the model is
  exceptional given the simplicity of the analysis performed, and inspires
  considerable confidence in the  fidelity of our mass estimates.}\label{fig:
  massfunc}
\end{figure*}



\section{Summary}\label{sec:summary}

In this work we present a new photometric dataset for astrophysics and
cosmology, KiDS+VIKING-450. The dataset builds on the optical dataset of 
KiDS-450 with the inclusion of 5-band near-IR data from the 
VIKING survey, reduced and analysed in a way entirely consistent 
with the optical dataset. 

We discuss the reduction of the VIKING dataset, and the derivation 
of relevant data products such as photometry. We demonstrate that 
the products derived are robust, consistent with, and superior to 
previous photometric estimates of sources from overlapping surveys 
such as 2MASS. 

Using our photometry, we derive new 9-band photometric redshifts for 
the full KV450 sample, and compare these new photo-$z$ to those 
presented previously in \cite{hildebrandt/etal:2017}. We find that the 
new photo-$z$ exhibit a reduced scatter in $\Delta z / (1+z)$ (especially 
at high photo-$z$; down by $\sim 40\%$ compared to the $\utoi$-only case),  
a lower overall bias (down $50\%$), and allow us to dramatically improve our 
ability to accurately estimate photo-$z$ beyond $z_{\rm B} = 0.9$, with the 
outlier rate reducing by over $40\%$. The improvement is sufficiently 
dramatic as to motivate the inclusion of a higher-redshift bin in KiDS 
cosmic shear studies using this dataset \citep{hildebrandt/etal:2018}, and to motivate us to explore 
whether our photo-$z$ alone are able to be used to constrain galaxy 
evolution parameters of interest (such as the stellar mass function) out 
to high redshift.  

Using the SED fitting code {\sc Le Phare}, we estimate stellar masses for 
all sources in the KV450 footprint. We compare these mass estimates to 
previous samples from GAMA \citep{wright/etal:2017} and G10-COSMOS 
\citep{andrews/etal:2017}, finding good agreement between the datasets. 
Our comparison to G10-COSMOS (a sample that matches the overall KV450 
dataset well) demonstrates negligible bias in our mass estimates 
($\mu_\Delta = 0.041$) and a scatter that is equivalent to that seen 
inherent to stellar mass estimates agnostic of changes to photometry and 
redshift \citep[$\sigma_\Delta = 0.202$;][]{wright/etal:2017,taylor/etal:2011}.
Furthermore, we demonstrate that the SED fits allow us to perform a 
high-fidelity star-galaxy separation, and thereby clean the full sample 
of contaminating sources. 

Using our mass estimates, we calculate the mass-completeness limit of 
the dataset, deriving an empirical mass limit that suggests the sample is 
volume complete above $M_\star\geq10^{10}M_\odot$ at $z_{\rm B}\leq 1$. 
We bin the data into eight volume complete samples spanning 
$0.1\leq z_{\rm B} \leq2$ and plot the resulting galaxy stellar mass functions for these 
bins. Comparing these bins to the evolutionary model of the GSMF from 
\cite{wright/etal:2018}, we find agreement in the range of 
$0.3\leq z_{\rm B}\leq2$. The lowest photo-$z$ bin shows considerable 
incompleteness at high-masses, which we attribute to our extraction pipeline 
being optimised for small-angular-size sources. In the regime where our 
pipeline is optimised, we demonstrate that we are able to reproduce the 
results of previous studies which utilised spectroscopic redshifts and/or significantly
more photometric data than we use here. Future KiDS+VIKING releases, containing three 
times the on-sky area utilised here, will further push the boundaries of studies that 
are possible with photometric-only data. 


\begin{acknowledgements}
This work is been supported by European Research Council Consolidator Grants: 770935 (AHW,HH) and 647112 (CH).  
HH is also supported by Emmy Noether (Hi 1495/2-1) and Heisenberg grants (Hi 1495/5-1) of the Deutsche
Forschungsgemeinschaft. KK acknowledges support by the Alexander von Humboldt Foundation. CH acknowledges support from
the Max Planck Society and the Alexander von Humboldt Foundation in the framework of the Max Planck-Humboldt Research
Award endowed by the Federal Ministry of Education and Research.  
AC acknowledges support from NASA grant 15-WFIRST15-0008. JTAdJ is supported by the Netherlands
Organisation for Scientific Research (NWO) through grant 621.016.402. ACE acknowledges support from STFC grant
ST/P00541/1. This work is supported by the Deutsche Forschungsgemeinschaft in the framework of the TR33 `The Dark
Universe'. Based on observations made with ESO Telescopes at the La Silla Paranal Observatory under programme IDs
179.A-2004, 177.A-3016, 177.A-3017, 177.A-3018, 298.A-5015. This work was performed in part at Aspen Center for Physics,
which is supported by National Science Foundation grant PHY-1607611.
\\
{\small \textit{Author Contributions:} 
All authors contributed to the development and writing of this paper. The
authorship list is given in three groups: the lead authors (AHW, HH, KK), 
followed by two alphabetical groups. The first alphabetical
group includes those who are key contributors to both the scientific analysis
and the data products. The second group covers those who have either made a
significant contribution to the data products or to the scientific analysis. }
\end{acknowledgements}

\bibpunct{(}{)}{;}{a}{}{,}
\bibliographystyle{aa}
\bibliography{KiDS-VIKING-450_data_paper}

\newcommand{\noopsort}[1]{}
\begin{thebibliography}{73}
\expandafter\ifx\csname natexlab\endcsname\relax\def\natexlab#1{#1}\fi

\bibitem[{Abbott {et~al.}(2016)}]{abbott/etal:2015}
Abbott, T. {et~al.} 2016, Phys. Rev., D94, 022001

\bibitem[{{Aihara} {et~al.}(2018){Aihara}, {Arimoto}, {Armstrong}, {Arnouts},
  {Bahcall}, {Bickerton}, {Bosch}, {Bundy}, {Capak}, {Chan}, {Chiba}, {Coupon},
  \& et~al.}]{aihara/etal:2018}
{Aihara}, H., {Arimoto}, N., {Armstrong}, R., {et~al.} 2018, \pasj, 70, S4

\bibitem[{{Amendola} {et~al.}(2018){Amendola}, {Appleby}, {Avgoustidis},
  {Bacon}, {Baker}, {Baldi}, {Bartolo}, {Blanchard}, {Bonvin}, {Borgani},
  {Branchini}, {Burrage}, {Camera}, {Carbone}, {Casarini}, {Cropper}, {de
  Rham}, {Dietrich}, {Di Porto}, {Durrer}, {Ealet}, {Ferreira}, {Finelli},
  {Garc{\'{\i}}a-Bellido}, {Giannantonio}, {Guzzo}, {Heavens}, {Heisenberg},
  {Heymans}, {Hoekstra}, {Hollenstein}, {Holmes}, {Hwang}, {Jahnke},
  {Kitching}, {Koivisto}, {Kunz}, {La Vacca}, {Linder}, {March}, {Marra},
  {Martins}, {Majerotto}, {Markovic}, {Marsh}, {Marulli}, {Massey}, {Mellier},
  {Montanari}, {Mota}, {Nunes}, {Percival}, {Pettorino}, {Porciani},
  {Quercellini}, {Read}, {Rinaldi}, {Sapone}, {Sawicki}, {Scaramella},
  {Skordis}, {Simpson}, {Taylor}, {Thomas}, {Trotta}, {Verde}, {Vernizzi},
  {Vollmer}, {Wang}, {Weller}, \& {Zlosnik}}]{amendola/etal:2018}
{Amendola}, L., {Appleby}, S., {Avgoustidis}, A., {et~al.} 2018, Living Reviews
  in Relativity, 21, 2

\bibitem[{{Andrews} {et~al.}(2017){Andrews}, {Driver}, {Davies}, {Kafle},
  {Robotham}, \& {Wright}}]{andrews/etal:2017}
{Andrews}, S.~K., {Driver}, S.~P., {Davies}, L.~J.~M., {et~al.} 2017, \mnras,
  464, 1569

\bibitem[{{Arnaboldi} {et~al.}(2007){Arnaboldi}, {Neeser}, {Parker}, {Rosati},
  {Lombardi}, {Dietrich}, \& {Hummel}}]{arnaboldi/etal:2007}
{Arnaboldi}, M., {Neeser}, M.~J., {Parker}, L.~C., {et~al.} 2007, The
  Messenger, 127

\bibitem[{{Arnouts} {et~al.}(1999){Arnouts}, {Cristiani}, {Moscardini},
  {Matarrese}, {Lucchin}, {Fontana}, \& {Giallongo}}]{arnouts/etal:1999}
{Arnouts}, S., {Cristiani}, S., {Moscardini}, L., {et~al.} 1999, \mnras, 310,
  540

\bibitem[{{Bacon} {et~al.}(2000){Bacon}, {Refregier}, \&
  {Ellis}}]{bacon/etal:2000}
{Bacon}, D.~J., {Refregier}, A.~R., \& {Ellis}, R.~S. 2000, \mnras, 318, 625

\bibitem[{{Balestra} {et~al.}(2010){Balestra}, {Mainieri}, {Popesso},
  {Dickinson}, {Nonino}, {Rosati}, {Teimoorinia}, {Vanzella}, {Cristiani},
  {Cesarsky}, {Fosbury}, {Kuntschner}, \& {Rettura}}]{balestra/etal:2010}
{Balestra}, I., {Mainieri}, V., {Popesso}, P., {et~al.} 2010, \aap, 512, A12

\bibitem[{{Banerji} {et~al.}(2015){Banerji}, {Jouvel}, {Lin}, {McMahon},
  {Lahav}, {Castander}, {Abdalla}, {Bertin}, {Bosman}, {Carnero}, {Kind}, {da
  Costa}, {Gerdes}, {Gschwend}, {Lima}, {Maia}, {Merson}, {Miller}, {Ogando},
  {Pellegrini}, {Reed}, {Saglia}, {S{\'a}nchez}, {Allam}, {Annis}, {Bernstein},
  {Bernstein}, {Bernstein}, {Capozzi}, {Childress}, {Cunha}, {Davis}, {DePoy},
  {Desai}, {Diehl}, {Doel}, {Findlay}, {Finley}, {Flaugher}, {Frieman},
  {Gaztanaga}, {Glazebrook}, {Gonz{\'a}lez-Fern{\'a}ndez}, {Gonzalez-Solares},
  {Honscheid}, {Irwin}, {Jarvis}, {Kim}, {Koposov}, {Kuehn}, {Kupcu-Yoldas},
  {Lagattuta}, {Lewis}, {Lidman}, {Makler}, {Marriner}, {Marshall}, {Miquel},
  {Mohr}, {Neilsen}, {Peoples}, {Sako}, {Sanchez}, {Scarpine}, {Schindler},
  {Schubnell}, {Sevilla}, {Sharp}, {Soares-Santos}, {Swanson}, {Tarle},
  {Thaler}, {Tucker}, {Uddin}, {Wechsler}, {Wester}, {Yuan}, \&
  {Zuntz}}]{banerji/etal:2015}
{Banerji}, M., {Jouvel}, S., {Lin}, H., {et~al.} 2015, \mnras, 446, 2523

\bibitem[{{Becker} {et~al.}(2016){Becker}, {Troxel}, {MacCrann}, {Krause},
  {Eifler}, {Friedrich}, {Nicola}, {Refregier}, {Amara}, {Bacon}, {Bernstein},
  {Bonnett}, {Bridle}, {Busha}, {Chang}, {Dodelson}, {Erickson}, {Evrard},
  {Frieman}, {Gaztanaga}, {Gruen}, {Hartley}, {Jain}, {Jarvis}, {Kacprzak},
  {Kirk}, {Kravtsov}, {Leistedt}, {Peiris}, {Rykoff}, {Sabiu}, {S{\'a}nchez},
  {Seo}, {Sheldon}, {Wechsler}, {Zuntz}, {Abbott}, {Abdalla}, {Allam},
  {Armstrong}, {Banerji}, {Bauer}, {Benoit-L{\'e}vy}, {Bertin}, {Brooks},
  {Buckley-Geer}, {Burke}, {Capozzi}, {Carnero Rosell}, {Carrasco Kind},
  {Carretero}, {Castander}, {Crocce}, {Cunha}, {D'Andrea}, {da Costa}, {DePoy},
  {Desai}, {Diehl}, {Dietrich}, {Doel}, {Fausti Neto}, {Fernandez}, {Finley},
  {Flaugher}, {Fosalba}, {Gerdes}, {Gruendl}, {Gutierrez}, {Honscheid},
  {James}, {Kuehn}, {Kuropatkin}, {Lahav}, {Li}, {Lima}, {Maia}, {March},
  {Martini}, {Melchior}, {Miller}, {Miquel}, {Mohr}, {Nichol}, {Nord},
  {Ogando}, {Plazas}, {Reil}, {Romer}, {Roodman}, {Sako}, {Sanchez},
  {Scarpine}, {Schubnell}, {Sevilla-Noarbe}, {Smith}, {Soares-Santos},
  {Sobreira}, {Suchyta}, {Swanson}, {Tarle}, {Thaler}, {Thomas}, {Vikram},
  {Walker}, \& {Dark Energy Survey Collaboration}}]{becker/etal:2016}
{Becker}, M.~R., {Troxel}, M.~A., {MacCrann}, N., {et~al.} 2016, \prd, 94,
  022002

\bibitem[{{Ben{\'{\i}}tez}(2000)}]{benitez:2000}
{Ben{\'{\i}}tez}, N. 2000, \apj, 536, 571

\bibitem[{{Bertin}(2010)}]{bertin:2010}
{Bertin}, E. 2010, {SWarp: Resampling and Co-adding FITS Images Together},
  Astrophysics Source Code Library

\bibitem[{{Bertin} \& {Arnouts}(1996)}]{bertin/arnouts:1996}
{Bertin}, E. \& {Arnouts}, S. 1996, \aaps, 117, 393

\bibitem[{{Bolzonella} {et~al.}(2000){Bolzonella}, {Miralles}, \&
  {Pell{\'o}}}]{bolzonella/etal:2000}
{Bolzonella}, M., {Miralles}, J.~M., \& {Pell{\'o}}, R. 2000, \aap, 363, 476

\bibitem[{{Brammer} {et~al.}(2008){Brammer}, {van Dokkum}, \&
  {Coppi}}]{brammer/etal:2008}
{Brammer}, G.~B., {van Dokkum}, P.~G., \& {Coppi}, P. 2008, \apj, 686, 1503

\bibitem[{{Bruzual} \& {Charlot}(2003)}]{bruzual/etal:2003}
{Bruzual}, G. \& {Charlot}, S. 2003, \mnras, 344, 1000

\bibitem[{{Buchs} {et~al.}(2019){Buchs}, {Davis}, {Gruen}, {DeRose}, {Alarcon},
  {Bernstein}, {S{\'a}nchez}, {Myles}, {Roodman}, {Allen}, {Amon}, \&
  et~al.}]{buchs/etal:2019}
{Buchs}, R., {Davis}, C., {Gruen}, D., {et~al.} 2019, arXiv e-prints,
  arXiv:1901.05005

\bibitem[{{Calzetti} {et~al.}(1994){Calzetti}, {Kinney}, \&
  {Storchi-Bergmann}}]{calzetti/etal:1994}
{Calzetti}, D., {Kinney}, A.~L., \& {Storchi-Bergmann}, T. 1994, \apj, 429, 582

\bibitem[{{Capak}(2004)}]{capak:2004}
{Capak}, P.~L. 2004, PhD thesis, UNIVERSITY OF HAWAI'I

\bibitem[{{Chabrier}(2003)}]{chabrier:2003}
{Chabrier}, G. 2003, \pasp, 115, 763

\bibitem[{{Cross} {et~al.}(2012){Cross}, {Collins}, {Mann}, {Read}, {Sutorius},
  {Blake}, {Holliman}, {Hambly}, {Emerson}, {Lawrence}, \&
  {Noddle}}]{cross/etal:2012}
{Cross}, N.~J.~G., {Collins}, R.~S., {Mann}, R.~G., {et~al.} 2012, \aap, 548,
  A119

\bibitem[{{Davidzon} {et~al.}(2017){Davidzon}, {Ilbert}, {Laigle}, {Coupon},
  {McCracken}, {Delvecchio}, {Masters}, {Capak}, {Hsieh}, {Le F{\`e}vre},
  {Tresse}, {Bethermin}, {Chang}, {Faisst}, {Le Floc'h}, {Steinhardt}, {Toft},
  {Aussel}, {Dubois}, {Hasinger}, {Salvato}, {Sanders}, {Scoville}, \&
  {Silverman}}]{davidzon/etal:2017}
{Davidzon}, I., {Ilbert}, O., {Laigle}, C., {et~al.} 2017, \aap, 605, A70

\bibitem[{{de Jong} {et~al.}(2017){de Jong}, {Verdois Kleijn}, {Erben},
  {Hildebrandt}, {Kuijken}, {Sikkema}, {Brescia}, {Bilicki}, {Napolitano},
  {Amaro}, {Begeman}, {Boxhoorn}, {Buddelmeijer}, {Cavuoti}, {Getman}, {Grado},
  {Helmich}, {Huang}, {Irisarri}, {La Barbera}, {Longo}, {McFarland},
  {Nakajima}, {Paolillo}, {Puddu}, {Radovich}, {Rifatto}, {Tortora},
  {Valentijn}, {Vellucci}, {Vriend}, {Amon}, {Blake}, {Choi}, {Conti}, {Gwyn},
  {Herbonnet}, {Heymans}, {Hoekstra}, {Klaes}, {Merten}, {Miller}, {Schneider},
  \& {Viola}}]{dejong/etal:2017}
{de Jong}, J. T.~A., {Verdois Kleijn}, G.~A., {Erben}, T., {et~al.} 2017, \aap,
  604, A134

\bibitem[{{\noopsort{De Jong}}{de Jong} {et~al.}(2015){\noopsort{De Jong}}{de
  Jong}, {Verdoes Kleijn}, {Boxhoorn}, {Buddelmeijer}, {Capaccioli}, {Getman},
  {Grado}, {Helmich}, {Huang}, {Irisarri}, {Kuijken}, {La Barbera},
  {McFarland}, {Napolitano}, {Radovich}, {Sikkema}, {Valentijn}, {Begeman},
  {Brescia}, {Cavuoti}, {Choi}, {Cordes}, {Covone}, {Dall'Ora}, {Hildebrandt},
  {Longo}, {Nakajima}, {Paolillo}, {Puddu}, {Rifatto}, {Tortora}, {van Uitert},
  {Buddendiek}, {Harnois-D{\'e}raps}, {Erben}, {Eriksen}, {Heymans},
  {Hoekstra}, {Joachimi}, {Kitching}, {Klaes}, {Koopmans}, {K{\"o}hlinger},
  {Roy}, {Sif{\'o}n}, {Schneider}, {Sutherland}, {Viola}, \&
  {Vriend}}]{dejong/etal:2015}
{\noopsort{De Jong}}{de Jong}, J.~T.~A., {Verdoes Kleijn}, G.~A., {Boxhoorn},
  D.~R., {et~al.} 2015, \aap, 582, A62

\bibitem[{{Driver} {et~al.}(2018){Driver}, {Andrews}, {da Cunha}, {Davies},
  {Lagos}, {Robotham}, {Vinsen}, {Wright}, {Alpaslan}, {Bland-Hawthorn},
  {Bourne}, {Brough}, {Bremer}, {Cluver}, {Colless}, {Conselice}, {Dunne},
  {Eales}, {Gomez}, {Holwerda}, {Hopkins}, {Kafle}, {Kelvin}, {Loveday},
  {Liske}, {Maddox}, {Phillipps}, {Pimbblet}, {Rowlands}, {Sansom}, {Taylor},
  {Wang}, \& {Wilkins}}]{driver/etal:2018}
{Driver}, S.~P., {Andrews}, S.~K., {da Cunha}, E., {et~al.} 2018, \mnras, 475,
  2891

\bibitem[{{Driver} {et~al.}(2011){Driver}, {Hill}, {Kelvin}, {Robotham},
  {Liske}, {Norberg}, {Baldry}, {Bamford}, {Hopkins}, {Loveday}, {Peacock},
  {Andrae}, {Bland-Hawthorn}, {Brough}, {Brown}, {Cameron}, {Ching}, {Colless},
  {Conselice}, {Croom}, {Cross}, {de Propris}, {Dye}, {Drinkwater}, {Ellis},
  {Graham}, {Grootes}, {Gunawardhana}, {Jones}, {van Kampen}, {Maraston},
  {Nichol}, {Parkinson}, {Phillipps}, {Pimbblet}, {Popescu}, {Prescott},
  {Roseboom}, {Sadler}, {Sansom}, {Sharp}, {Smith}, {Taylor}, {Thomas},
  {Tuffs}, {Wijesinghe}, {Dunne}, {Frenk}, {Jarvis}, {Madore}, {Meyer},
  {Seibert}, {Staveley-Smith}, {Sutherland}, \& {Warren}}]{driver/etal:2011}
{Driver}, S.~P., {Hill}, D.~T., {Kelvin}, L.~S., {et~al.} 2011, \mnras, 413,
  971

\bibitem[{{Driver} {et~al.}(2016){Driver}, {Wright}, {Andrews}, {Davies},
  {Kafle}, {Lange}, {Moffett}, {Mannering}, {Robotham}, {Vinsen}, {Alpaslan},
  {Andrae}, {Baldry}, {Bauer}, {Bamford}, {Bland-Hawthorn}, {Bourne}, {Brough},
  {Brown}, {Cluver}, {Croom}, {Colless}, {Conselice}, {da Cunha}, {De Propris},
  {Drinkwater}, {Dunne}, {Eales}, {Edge}, {Frenk}, {Graham}, {Grootes},
  {Holwerda}, {Hopkins}, {Ibar}, {van Kampen}, {Kelvin}, {Jarrett}, {Jones},
  {Lara-Lopez}, {Liske}, {Lopez-Sanchez}, {Loveday}, {Maddox}, {Madore},
  {Mahajan}, {Meyer}, {Norberg}, {Penny}, {Phillipps}, {Popescu}, {Tuffs},
  {Peacock}, {Pimbblet}, {Prescott}, {Rowlands}, {Sansom}, {Seibert}, {Smith},
  {Sutherland}, {Taylor}, {Valiante}, {Vazquez-Mata}, {Wang}, {Wilkins}, \&
  {Williams}}]{driver/etal:2016}
{Driver}, S.~P., {Wright}, A.~H., {Andrews}, S.~K., {et~al.} 2016, \mnras, 455,
  3911

\bibitem[{{Edge} {et~al.}(2013){Edge}, {Sutherland}, {Kuijken}, {Driver},
  {McMahon}, {Eales}, \& {Emerson}}]{edge/etal:2013}
{Edge}, A., {Sutherland}, W., {Kuijken}, K., {et~al.} 2013, The Messenger, 154,
  32

\bibitem[{{Erben} {et~al.}(2005){Erben}, {Schirmer}, {Dietrich}, {Cordes},
  {Haberzettl}, {Hetterscheidt}, {Hildebrandt}, {Schmithuesen}, {Schneider},
  {Simon}, {Deul}, {Hook}, {Kaiser}, {Radovich}, {Benoist}, {Nonino}, {Olsen},
  {Prandoni}, {Wichmann}, {Zaggia}, {Bomans}, {Dettmar}, \&
  {Miralles}}]{erben/etal:2005}
{Erben}, T., {Schirmer}, M., {Dietrich}, J.~P., {et~al.} 2005, Astronomische
  Nachrichten, 326, 432

\bibitem[{{Gaia Collaboration} {et~al.}(2018){Gaia Collaboration}, {Brown},
  {Vallenari}, {Prusti}, {de Bruijne}, {Babusiaux}, {Bailer-Jones}, {Biermann},
  {Evans}, {Eyer}, {Jansen}, {Jordi}, {Klioner}, \& et~al.}]{GaiaDR2}
{Gaia Collaboration}, {Brown}, A.~G.~A., {Vallenari}, A., {et~al.} 2018, \aap,
  616, A1

\bibitem[{{Gaia Collaboration} {et~al.}(2016){Gaia Collaboration}, {Brown},
  {Vallenari}, {Prusti}, {de Bruijne}, {Mignard}, {Drimmel}, {Babusiaux},
  {Bailer-Jones}, {Bastian}, \& et~al.}]{GaiaDR1}
{Gaia Collaboration}, {Brown}, A.~G.~A., {Vallenari}, A., {et~al.} 2016, \aap,
  595, A2

\bibitem[{{Gonz{\'a}lez-Fern{\'a}ndez}
  {et~al.}(2018){Gonz{\'a}lez-Fern{\'a}ndez}, {Hodgkin}, {Irwin},
  {Gonz{\'a}lez-Solares}, {Koposov}, {Lewis}, {Emerson}, {Hewett}, {Yolda{\c
  s}}, \& {Riello}}]{gonzalez-fernandez/etal:2018}
{Gonz{\'a}lez-Fern{\'a}ndez}, C., {Hodgkin}, S.~T., {Irwin}, M.~J., {et~al.}
  2018, \mnras, 474, 5459

\bibitem[{{Hambly} {et~al.}(2008){Hambly}, {Collins}, {Cross}, {Mann}, {Read},
  {Sutorius}, {Bond}, {Bryant}, {Emerson}, {Lawrence}, {Rimoldini}, {Stewart},
  {Williams}, {Adamson}, {Hirst}, {Dye}, \& {Warren}}]{hambly/etal:2008}
{Hambly}, N.~C., {Collins}, R.~S., {Cross}, N.~J.~G., {et~al.} 2008, \mnras,
  384, 637

\bibitem[{{High} {et~al.}(2009){High}, {Stubbs}, {Rest}, {Stalder}, \&
  {Challis}}]{high/etal:2009}
{High}, F.~W., {Stubbs}, C.~W., {Rest}, A., {Stalder}, B., \& {Challis}, P.
  2009, \aj, 138, 110

\bibitem[{{Hikage} {et~al.}(2018){Hikage}, {Oguri}, {Hamana}, {More},
  {Mandelbaum}, {Takada}, {K{\"o}hlinger}, {Miyatake}, {Nishizawa}, {Aihara},
  {Armstrong}, {Bosch}, {Coupon}, {Ducout}, {Hsieh}, {Komiyama}, {Lanusse},
  {Leauthaud}, {Medezinski}, {Mineo}, {Miyazaki}, {Murata}, {Murayama},
  {Shirasaki}, {Sif{\'o}n}, {Simet}, {Speagle}, {Spergel}, {Strauss},
  {Sugiyama}, {Tanaka}, \& {Wang}}]{hikage/etal:2018}
{Hikage}, C., {Oguri}, M., {Hamana}, T., {et~al.} 2018, ArXiv e-prints
  [\eprint[arXiv]{1809.09148}]

\bibitem[{{Hildebrandt} {et~al.}(2010){Hildebrandt}, {Arnouts}, {Capak},
  {Moustakas}, {Wolf}, {Abdalla}, {Assef}, {Banerji}, {Ben{\'{\i}}tez},
  {Brammer}, {Budav{\'a}ri}, {Carliles}, {Coe}, {Dahlen}, {Feldmann}, {Gerdes},
  {Gillis}, {Ilbert}, {Kotulla}, {Lahav}, {Li}, {Miralles}, {Purger},
  {Schmidt}, \& {Singal}}]{hildebrandt/etal:2010}
{Hildebrandt}, H., {Arnouts}, S., {Capak}, P., {et~al.} 2010, \aap, 523, A31

\bibitem[{{Hildebrandt} {et~al.}(2016){Hildebrandt}, {Choi}, {Heymans},
  {Blake}, {Erben}, {Miller}, {Nakajima}, {van Waerbeke}, {Viola},
  {Buddendiek}, {Harnois-D{\'e}raps}, {Hojjati}, {Joachimi}, {Joudaki},
  {Kitching}, {Wolf}, {Gwyn}, {Johnson}, {Kuijken}, {Sheikhbahaee}, {Tudorica},
  \& {Yee}}]{hildebrandt/etal:2016}
{Hildebrandt}, H., {Choi}, A., {Heymans}, C., {et~al.} 2016, \mnras, 463, 635

\bibitem[{{Hildebrandt} {et~al.}(2018){Hildebrandt}, {K{\"o}hlinger}, {van den
  Busch}, {Joachimi}, {Heymans}, {Kannawadi}, {Wright}, {Asgari}, {Blake},
  {Hoekstra}, {Joudaki}, {Kuijken}, {Miller}, {Morrison}, {Tr{\"o}ster},
  {Amon}, {Archidiacono}, {Brieden}, {Choi}, {de Jong}, {Erben}, {Giblin},
  {Mead}, {Peacock}, {Radovich}, {Schneider}, {Sif{\'o}n}, \&
  {Tewes}}]{hildebrandt/etal:2018}
{Hildebrandt}, H., {K{\"o}hlinger}, F., {van den Busch}, J.~L., {et~al.} 2018,
  arXiv e-prints, arXiv:1812.06076

\bibitem[{{Hildebrandt} {et~al.}(2017){Hildebrandt}, {Viola}, {Heymans},
  {Joudaki}, {Kuijken}, {Blake}, {Erben}, {Joachimi}, {Klaes}, {Miller},
  {Morrison}, {Nakajima}, {Verdoes Kleijn}, {Amon}, {Choi}, {Covone}, {de
  Jong}, {Dvornik}, {Fenech Conti}, {Grado}, {Harnois-D{\'e}raps}, {Herbonnet},
  {Hoekstra}, {K{\"o}hlinger}, {McFarland}, {Mead}, {Merten}, {Napolitano},
  {Peacock}, {Radovich}, {Schneider}, {Simon}, {Valentijn}, {van den Busch},
  {van Uitert}, \& {Van Waerbeke}}]{hildebrandt/etal:2017}
{Hildebrandt}, H., {Viola}, M., {Heymans}, C., {et~al.} 2017, \mnras, 465, 1454

\bibitem[{{Ilbert} {et~al.}(2006){Ilbert}, {Arnouts}, {McCracken},
  {Bolzonella}, {Bertin}, {Le F{\`e}vre}, {Mellier}, {Zamorani}, {Pell{\`o}},
  {Iovino}, {Tresse}, {Le Brun}, {Bottini}, {Garilli}, {Maccagni}, {Picat},
  {Scaramella}, {Scodeggio}, {Vettolani}, {Zanichelli}, {Adami}, {Bardelli},
  {Cappi}, {Charlot}, {Ciliegi}, {Contini}, {Cucciati}, {Foucaud}, {Franzetti},
  {Gavignaud}, {Guzzo}, {Marano}, {Marinoni}, {Mazure}, {Meneux}, {Merighi},
  {Paltani}, {Pollo}, {Pozzetti}, {Radovich}, {Zucca}, {Bondi}, {Bongiorno},
  {Busarello}, {de La Torre}, {Gregorini}, {Lamareille}, {Mathez}, {Merluzzi},
  {Ripepi}, {Rizzo}, \& {Vergani}}]{ilbert/etal:2006}
{Ilbert}, O., {Arnouts}, S., {McCracken}, H.~J., {et~al.} 2006, \aap, 457, 841

\bibitem[{{Ilbert} {et~al.}(2009){Ilbert}, {Capak}, {Salvato}, {Aussel},
  {McCracken}, {Sanders}, {Scoville}, {Kartaltepe}, {Arnouts}, {Le Floc'h},
  {Mobasher}, {Taniguchi}, {Lamareille}, {Leauthaud}, {Sasaki}, {Thompson},
  {Zamojski}, {Zamorani}, {Bardelli}, {Bolzonella}, {Bongiorno}, {Brusa},
  {Caputi}, {Carollo}, {Contini}, {Cook}, {Coppa}, {Cucciati}, {de la Torre},
  {de Ravel}, {Franzetti}, {Garilli}, {Hasinger}, {Iovino}, {Kampczyk},
  {Kneib}, {Knobel}, {Kovac}, {Le Borgne}, {Le Brun}, {Le F{\`e}vre}, {Lilly},
  {Looper}, {Maier}, {Mainieri}, {Mellier}, {Mignoli}, {Murayama}, {Pell{\`o}},
  {Peng}, {P{\'e}rez-Montero}, {Renzini}, {Ricciardelli}, {Schiminovich},
  {Scodeggio}, {Shioya}, {Silverman}, {Surace}, {Tanaka}, {Tasca}, {Tresse},
  {Vergani}, \& {Zucca}}]{ilbert/etal:2009}
{Ilbert}, O., {Capak}, P., {Salvato}, M., {et~al.} 2009, \apj, 690, 1236

\bibitem[{{Irwin} {et~al.}(2004){Irwin}, {Lewis}, {Hodgkin}, {Bunclark},
  {Evans}, {McMahon}, {Emerson}, {Stewart}, \& {Beard}}]{Irwin/etal:2004}
{Irwin}, M.~J., {Lewis}, J., {Hodgkin}, S., {et~al.} 2004, in \procspie, Vol.
  5493, Optimizing Scientific Return for Astronomy through Information
  Technologies, ed. P.~J. {Quinn} \& A.~{Bridger}, 411--422

\bibitem[{{Jee} {et~al.}(2016){Jee}, {Tyson}, {Hilbert}, {Schneider},
  {Schmidt}, \& {Wittman}}]{jee/etal:2016}
{Jee}, M.~J., {Tyson}, J.~A., {Hilbert}, S., {et~al.} 2016, \apj, 824, 77

\bibitem[{{Joachimi} {et~al.}(2015){Joachimi}, {Cacciato}, {Kitching},
  {Leonard}, {Mandelbaum}, {Sch{\"a}fer}, {Sif{\'o}n}, {Hoekstra}, {Kiessling},
  {Kirk}, \& {Rassat}}]{joachimi/etal:2015}
{Joachimi}, B., {Cacciato}, M., {Kitching}, T.~D., {et~al.} 2015, \ssr, 193, 1

\bibitem[{{Kafle} {et~al.}(2018){Kafle}, {Robotham}, {Driver}, {Deeley},
  {Norberg}, {Drinkwater}, \& {Davies}}]{kafle/etal:2018}
{Kafle}, P.~R., {Robotham}, A.~S.~G., {Driver}, S.~P., {et~al.} 2018, \mnras,
  479, 3746

\bibitem[{{Kron}(1980)}]{kron:1980}
{Kron}, R.~G. 1980, \apjs, 43, 305

\bibitem[{{Kuij\-ken}(2008)}]{kuijken:2008}
{Kuij\-ken}, K. 2008, \aap, 482, 1053

\bibitem[{{Kuij\-ken} {et~al.}(2015){Kuij\-ken}, {Heymans}, {Hildebrandt},
  {Nakajima}, {Erben}, {de Jong}, {Viola}, {Choi}, {Hoekstra}, {Miller}, {van
  Uitert}, {Amon}, {Blake}, {Brouwer}, {Buddendiek}, {Conti}, {Eriksen},
  {Grado}, {Harnois-D{\'e}raps}, {Helmich}, {Herbonnet}, {Irisarri},
  {Kitching}, {Klaes}, {La Barbera}, {Napolitano}, {Radovich}, {Schneider},
  {Sif{\'o}n}, {Sikkema}, {Simon}, {Tudorica}, {Valentijn}, {Verdoes Kleijn},
  \& {van Waerbeke}}]{kuijken/etal:2015}
{Kuij\-ken}, K., {Heymans}, C., {Hildebrandt}, H., {et~al.} 2015, \mnras, 454,
  3500

\bibitem[{{Laigle} {et~al.}(2019){Laigle}, {Davidzon}, {Ilbert}, {Devriendt},
  {Kashino}, {Pichon}, {Capak}, {Arnouts}, {de la Torre}, {Dubois},
  {Gozaliasl}, {Le Borgne}, {Lilly}, {McCracken}, {Salvato}, \&
  {Slyz}}]{laigle/etal:2019}
{Laigle}, C., {Davidzon}, I., {Ilbert}, O., {et~al.} 2019, \mnras, 486, 5104

\bibitem[{{Le F{\`e}vre} {et~al.}(2013){Le F{\`e}vre}, {Cassata}, {Cucciati},
  {Garilli}, {Ilbert}, {Le Brun}, {Maccagni}, {Moreau}, {Scodeggio}, {Tresse},
  {Zamorani}, {Adami}, {Arnouts}, {Bardelli}, {Bolzonella}, {Bondi},
  {Bongiorno}, {Bottini}, {Cappi}, {Charlot}, {Ciliegi}, {Contini}, {de la
  Torre}, {Foucaud}, {Franzetti}, {Gavignaud}, {Guzzo}, {Iovino}, {Lemaux},
  {L{\'o}pez-Sanjuan}, {McCracken}, {Marano}, {Marinoni}, {Mazure}, {Mellier},
  {Merighi}, {Merluzzi}, {Paltani}, {Pell{\`o}}, {Pollo}, {Pozzetti},
  {Scaramella}, {Tasca}, {Vergani}, {Vettolani}, {Zanichelli}, \&
  {Zucca}}]{lefevre/etal:2013}
{Le F{\`e}vre}, O., {Cassata}, P., {Cucciati}, O., {et~al.} 2013, \aap, 559,
  A14

\bibitem[{{Lewis} {et~al.}(2010){Lewis}, {Irwin}, \&
  {Bunclark}}]{lewis/etal:2010}
{Lewis}, J.~R., {Irwin}, M., \& {Bunclark}, P. 2010, in Astronomical Society of
  the Pacific Conference Series, Vol. 434, Astronomical Data Analysis Software
  and Systems XIX, ed. Y.~{Mizumoto}, K.-I. {Morita}, \& M.~{Ohishi}, 91

\bibitem[{{Lilly} {et~al.}(2009){Lilly}, {Le Brun}, {Maier}, {Mainieri},
  {Mignoli}, {Scodeggio}, {Zamorani}, {Carollo}, {Contini}, {Kneib}, {Le
  F{\`e}vre}, {Renzini}, {Bardelli}, {Bolzonella}, {Bongiorno}, {Caputi},
  {Coppa}, {Cucciati}, {de la Torre}, {de Ravel}, {Franzetti}, {Garilli},
  {Iovino}, {Kampczyk}, {Kovac}, {Knobel}, {Lamareille}, {Le Borgne}, {Pello},
  {Peng}, {P{\'e}rez-Montero}, {Ricciardelli}, {Silverman}, {Tanaka}, {Tasca},
  {Tresse}, {Vergani}, {Zucca}, {Ilbert}, {Salvato}, {Oesch}, {Abbas},
  {Bottini}, {Capak}, {Cappi}, {Cassata}, {Cimatti}, {Elvis}, {Fumana},
  {Guzzo}, {Hasinger}, {Koekemoer}, {Leauthaud}, {Maccagni}, {Marinoni},
  {McCracken}, {Memeo}, {Meneux}, {Porciani}, {Pozzetti}, {Sanders},
  {Scaramella}, {Scarlata}, {Scoville}, {Shopbell}, \&
  {Taniguchi}}]{lilly/etal:2009}
{Lilly}, S.~J., {Le Brun}, V., {Maier}, C., {et~al.} 2009, \apjs, 184, 218

\bibitem[{{Mandelbaum}(2018)}]{mandelbaum:2018}
{Mandelbaum}, R. 2018, Annual Review of Astronomy and Astrophysics, 56, 393

\bibitem[{{Massey} {et~al.}(2007){Massey}, {Heymans}, {Berg{\'e}}, {Bernstein},
  {Bridle}, {Clowe}, {Dahle}, {Ellis}, {Erben}, {Hetterscheidt}, {High},
  {Hirata}, {Hoekstra}, {Hudelot}, {Jarvis}, {Johnston}, {Kuijken},
  {Margoniner}, {Mandelbaum}, {Mellier}, {Nakajima}, {Paulin-Henriksson},
  {Peeples}, {Roat}, {Refregier}, {Rhodes}, {Schrabback}, {Schirmer}, {Seljak},
  {Semboloni}, \& {van Waerbeke}}]{massey/etal:2007a}
{Massey}, R., {Heymans}, C., {Berg{\'e}}, J., {et~al.} 2007, \mnras, 376, 13

\bibitem[{{Newman} {et~al.}(2013){Newman}, {Cooper}, {Davis}, {Faber}, {Coil},
  {Guhathakurta}, {Koo}, {Phillips}, {Conroy}, {Dutton}, {Finkbeiner}, {Gerke},
  {Rosario}, {Weiner}, {Willmer}, {Yan}, {Harker}, {Kassin}, {Konidaris},
  {Lai}, {Madgwick}, {Noeske}, {Wirth}, {Connolly}, {Kaiser}, {Kirby},
  {Lemaux}, {Lin}, {Lotz}, {Luppino}, {Marinoni}, {Matthews}, {Metevier}, \&
  {Schiavon}}]{newman/etal:2013}
{Newman}, J.~A., {Cooper}, M.~C., {Davis}, M., {et~al.} 2013, \apjs, 208, 5

\bibitem[{{Planck Collaboration} {et~al.}(2018){Planck Collaboration},
  {Aghanim}, {Akrami}, {Ashdown}, {Aumont}, {Baccigalupi}, {Ballardini},
  {Banday}, {Barreiro}, {Bartolo}, {Basak}, {Battye}, {Benabed}, {Bernard},
  {Bersanelli}, {Bielewicz}, {Bock}, {Bond}, {Borrill}, {Bouchet}, {Boulanger},
  {Bucher}, {Burigana}, {Butler}, {Calabrese}, {Cardoso}, {Carron},
  {Challinor}, {Chiang}, {Chluba}, {Colombo}, {Combet}, {Contreras}, {Crill},
  {Cuttaia}, {de Bernardis}, {de Zotti}, {Delabrouille}, {Delouis}, {Di
  Valentino}, {Diego}, {Dor{\'e}}, {Douspis}, {Ducout}, {Dupac}, {Dusini},
  {Efstathiou}, {Elsner}, {En{\ss}lin}, {Eriksen}, {Fantaye}, {Farhang},
  {Fergusson}, {Fernandez-Cobos}, {Finelli}, {Forastieri}, {Frailis},
  {Franceschi}, {Frolov}, {Galeotta}, {Galli}, {Ganga}, {G{\'e}nova-Santos},
  {Gerbino}, {Ghosh}, {Gonz{\'a}lez-Nuevo}, {G{\'o}rski}, {Gratton},
  {Gruppuso}, {Gudmundsson}, {Hamann}, {Handley}, {Herranz}, {Hivon}, {Huang},
  {Jaffe}, {Jones}, {Karakci}, {Keih{\"a}nen}, {Keskitalo}, {Kiiveri}, {Kim},
  {Kisner}, {Knox}, {Krachmalnicoff}, {Kunz}, {Kurki-Suonio}, {Lagache},
  {Lamarre}, {Lasenby}, {Lattanzi}, {Lawrence}, {Le Jeune}, {Lemos},
  {Lesgourgues}, {Levrier}, {Lewis}, {Liguori}, {Lilje}, {Lilley}, {Lindholm},
  {L{\'o}pez-Caniego}, {Lubin}, {Ma}, {Mac{\'{\i}}as-P{\'e}rez}, {Maggio},
  {Maino}, {Mandolesi}, {Mangilli}, {Marcos-Caballero}, {Maris}, {Martin},
  {Martinelli}, {Mart{\'{\i}}nez-Gonz{\'a}lez}, {Matarrese}, {Mauri}, {McEwen},
  {Meinhold}, {Melchiorri}, {Mennella}, {Migliaccio}, {Millea}, {Mitra},
  {Miville-Desch{\^e}nes}, {Molinari}, {Montier}, {Morgante}, {Moss}, {Natoli},
  {N{\o}rgaard-Nielsen}, {Pagano}, {Paoletti}, {Partridge}, {Patanchon},
  {Peiris}, {Perrotta}, {Pettorino}, {Piacentini}, {Polastri}, {Polenta},
  {Puget}, {Rachen}, {Reinecke}, {Remazeilles}, {Renzi}, {Rocha}, {Rosset},
  {Roudier}, {Rubi{\~n}o-Mart{\'{\i}}n}, {Ruiz-Granados}, {Salvati}, {Sandri},
  {Savelainen}, {Scott}, {Shellard}, {Sirignano}, {Sirri}, {Spencer},
  {Sunyaev}, {Suur-Uski}, {Tauber}, {Tavagnacco}, {Tenti}, {Toffolatti},
  {Tomasi}, {Trombetti}, {Valenziano}, {Valiviita}, {Van Tent}, {Vibert},
  {Vielva}, {Villa}, {Vittorio}, {Wandelt}, {Wehus}, {White}, {White},
  {Zacchei}, \& {Zonca}}]{planck/cosmo:2018}
{Planck Collaboration}, {Aghanim}, N., {Akrami}, Y., {et~al.} 2018, ArXiv
  e-prints [\eprint[arXiv]{1807.06209}]

\bibitem[{{Popesso} {et~al.}(2009){Popesso}, {Dickinson}, {Nonino}, {Vanzella},
  {Daddi}, {Fosbury}, {Kuntschner}, {Mainieri}, {Cristiani}, {Cesarsky},
  {Giavalisco}, {Renzini}, \& {GOODS Team}}]{popesso/etal:2009}
{Popesso}, P., {Dickinson}, M., {Nonino}, M., {et~al.} 2009, \aap, 494, 443

\bibitem[{{Raichoor} {et~al.}(2014){Raichoor}, {Mei}, {Erben}, {Hildebrandt},
  {Huertas-Company}, {Ilbert}, {Licitra}, {Ball}, {Boissier}, {Boselli},
  {Chen}, {C{\^o}t{\'e}}, {Cuillandre}, {Duc}, {Durrell}, {Ferrarese},
  {Guhathakurta}, {Gwyn}, {Kavelaars}, {Lan{\c c}on}, {Liu}, {MacArthur},
  {Muller}, {Mu{\~n}oz}, {Peng}, {Puzia}, {Sawicki}, {Toloba}, {Van Waerbeke},
  {Woods}, \& {Zhang}}]{raichoor/etal:2014}
{Raichoor}, A., {Mei}, S., {Erben}, T., {et~al.} 2014, \apj, 797, 102

\bibitem[{{Refregier}(2003)}]{refregier:2003}
{Refregier}, A. 2003, \mnras, 338, 35

\bibitem[{{Rhodes} {et~al.}(2001){Rhodes}, {Refregier}, \&
  {Groth}}]{rhodes/etal:2001}
{Rhodes}, J., {Refregier}, A., \& {Groth}, E.~J. 2001, \apjl, 552, L85

\bibitem[{{Schirmer}(2013)}]{schirmer:2013}
{Schirmer}, M. 2013, \apjs, 209, 21

\bibitem[{{Schlegel} {et~al.}(1998){Schlegel}, {Finkbeiner}, \&
  {Davis}}]{schlegel/etal:1998}
{Schlegel}, D.~J., {Finkbeiner}, D.~P., \& {Davis}, M. 1998, \apj, 500, 525

\bibitem[{{Taylor} {et~al.}(2011){Taylor}, {Hopkins}, {Baldry}, {Brown},
  {Driver}, {Kelvin}, {Hill}, {Robotham}, {Bland-Hawthorn}, {Jones}, {Sharp},
  {Thomas}, {Liske}, {Loveday}, {Norberg}, {Peacock}, {Bamford}, {Brough},
  {Colless}, {Cameron}, {Conselice}, {Croom}, {Frenk}, {Gunawardhana},
  {Kuijken}, {Nichol}, {Parkinson}, {Phillipps}, {Pimbblet}, {Popescu},
  {Prescott}, {Sutherland}, {Tuffs}, {van Kampen}, \&
  {Wijesinghe}}]{taylor/etal:2011}
{Taylor}, E.~N., {Hopkins}, A.~M., {Baldry}, I.~K., {et~al.} 2011, \mnras, 418,
  1587

\bibitem[{{Tortora} {et~al.}(2018){Tortora}, {Napolitano}, {Spavone}, {La
  Barbera}, {D'Ago}, {Spiniello}, {Kuijken}, {Roy}, {Raj}, {Cavuoti},
  {Brescia}, {Longo}, {Pota}, {Petrillo}, {Radovich}, {Getman}, {Koopmans},
  {Trujillo}, {Verdoes Kleijn}, {Capaccioli}, {Grado}, {Covone},
  {Scognamiglio}, {Blake}, {Glazebrook}, {Joudaki}, {Lidman}, \&
  {Wolf}}]{tortora/etal:2018}
{Tortora}, C., {Napolitano}, N.~R., {Spavone}, M., {et~al.} 2018, \mnras, 481,
  4728

\bibitem[{{Troxel} {et~al.}(2018){Troxel}, {Krause}, {Chang}, {Eifler},
  {Friedrich}, {Gruen}, {MacCrann}, {Chen}, {Davis}, {DeRose}, {Dodelson},
  {Gatti}, {Hoyle}, {Huterer}, {Jarvis}, {Lacasa}, {Lemos}, {Peiris}, {Prat},
  {Samuroff}, {S{\'a}nchez}, {Sheldon}, {Vielzeuf}, {Wang}, {Zuntz}, {Lahav},
  {Abdalla}, {Allam}, {Annis}, {Avila}, {Bertin}, {Brooks}, {Burke}, {Carnero
  Rosell}, {Carrasco Kind}, {Carretero}, {Crocce}, {Cunha}, {D'Andrea}, {da
  Costa}, {De Vicente}, {Diehl}, {Doel}, {Evrard}, {Flaugher}, {Fosalba},
  {Frieman}, {Garc{\'\i}a-Bellido}, {Gaztanaga}, {Gerdes}, {Gruendl},
  {Gschwend}, {Gutierrez}, {Hartley}, {Hollowood}, {Honscheid}, {James},
  {Kirk}, {Kuehn}, {Kuropatkin}, {Li}, {Lima}, {March}, {Menanteau}, {Miquel},
  {Mohr}, {Ogando}, {Plazas}, {Roodman}, {Sanchez}, {Scarpine}, {Schindler},
  {Sevilla-Noarbe}, {Smith}, {Soares-Santos}, {Sobreira}, {Suchyta}, {Swanson},
  {Thomas}, {Walker}, \& {Wechsler}}]{troxel/etal:2018}
{Troxel}, M.~A., {Krause}, E., {Chang}, C., {et~al.} 2018, \mnras, 479, 4998

\bibitem[{{Valentijn} {et~al.}(2007){Valentijn}, {McFarland}, {Snigula},
  {Begeman}, {Boxhoorn}, {Rengelink}, {Helmich}, {Heraudeau}, {Verdoes Kleijn},
  {Vermeij}, {Vriend}, {Tempelaar}, {Deul}, {Kuijken}, {Capaccioli},
  {Silvotti}, {Bender}, {Neeser}, {Saglia}, {Bertin}, \&
  {Mellier}}]{valentijn/etal:2007}
{Valentijn}, E.~A., {McFarland}, J.~P., {Snigula}, J., {et~al.} 2007, in
  Astronomical Society of the Pacific Conference Series, Vol. 376, Astronomical
  Data Analysis Software and Systems XVI, ed. R.~A. {Shaw}, F.~{Hill}, \& D.~J.
  {Bell}, 491

\bibitem[{{Van Waerbeke} {et~al.}(2000){Van Waerbeke}, {Mellier}, {Erben},
  {Cuilland re}, {Bernardeau}, {Maoli}, {Bertin}, {McCracken}, {Le F{\`e}vre},
  {Fort}, {Dantel-Fort}, {Jain}, \& {Schneider}}]{vanwaerbeke/etal:2000}
{Van Waerbeke}, L., {Mellier}, Y., {Erben}, T., {et~al.} 2000, \aap, 358, 30

\bibitem[{{Vanzella} {et~al.}(2008){Vanzella}, {Cristiani}, {Dickinson},
  {Giavalisco}, {Kuntschner}, {Haase}, {Nonino}, {Rosati}, {Cesarsky},
  {Ferguson}, {Fosbury}, {Grazian}, {Moustakas}, {Rettura}, {Popesso},
  {Renzini}, {Stern}, \& {GOODS Team}}]{vanzella/etal:2008}
{Vanzella}, E., {Cristiani}, S., {Dickinson}, M., {et~al.} 2008, \aap, 478, 83

\bibitem[{{Venemans} {et~al.}(2015){Venemans}, {Verdoes Kleijn}, {Mwebaze},
  {Valentijn}, {Ba{\~n}ados}, {Decarli}, {de Jong}, {Findlay}, {Kuijken},
  {Barbera}, {McFarland}, {McMahon}, {Napolitano}, {Sikkema}, \&
  {Sutherland}}]{venemans/etal:2015}
{Venemans}, B.~P., {Verdoes Kleijn}, G.~A., {Mwebaze}, J., {et~al.} 2015,
  \mnras, 453, 2259

\bibitem[{{Wittman} {et~al.}(2000){Wittman}, {Tyson}, {Kirkman},
  {Dell'Antonio}, \& {Bernstein}}]{wittman/etal:2000}
{Wittman}, D.~M., {Tyson}, J.~A., {Kirkman}, D., {Dell'Antonio}, I., \&
  {Bernstein}, G. 2000, \nat, 405, 143

\bibitem[{{Wright} {et~al.}(2018){Wright}, {Driver}, \&
  {Robotham}}]{wright/etal:2018}
{Wright}, A.~H., {Driver}, S.~P., \& {Robotham}, A.~S.~G. 2018, \mnras, 480,
  3491

\bibitem[{{Wright} {et~al.}(2016){Wright}, {Robotham}, {Bourne}, {Driver},
  {Dunne}, {Maddox}, {Alpaslan}, {Andrews}, {Bauer}, {Bland-Hawthorn},
  {Brough}, {Brown}, {Clarke}, {Cluver}, {Davies}, {Grootes}, {Holwerda},
  {Hopkins}, {Jarrett}, {Kafle}, {Lange}, {Liske}, {Loveday}, {Moffett},
  {Norberg}, {Popescu}, {Smith}, {Taylor}, {Tuffs}, {Wang}, \&
  {Wilkins}}]{wright/etal:2016}
{Wright}, A.~H., {Robotham}, A.~S.~G., {Bourne}, N., {et~al.} 2016, \mnras,
  460, 765

\bibitem[{{Wright} {et~al.}(2017){Wright}, {Robotham}, {Driver}, {Alpaslan},
  {Andrews}, {Baldry}, {Bland-Hawthorn}, {Brough}, {Brown}, {Colless}, {da
  Cunha}, {Davies}, {Graham}, {Holwerda}, {Hopkins}, {Kafle}, {Kelvin},
  {Loveday}, {Maddox}, {Meyer}, {Moffett}, {Norberg}, {Phillipps}, {Rowlands},
  {Taylor}, {Wang}, \& {Wilkins}}]{wright/etal:2017}
{Wright}, A.~H., {Robotham}, A.~S.~G., {Driver}, S.~P., {et~al.} 2017, \mnras,
  470, 283

\end{thebibliography}

\clearpage
\appendix 


%
%
%

\end{document}